\documentclass[twocolumn,superscriptaddress,amsmath,amssymb,floatfix,aps,notitlepage,nokeys,pre]{revtex4}
\usepackage{physics}
\usepackage{mathrsfs, hyperref}
\usepackage{amssymb,amsbsy,amsmath,latexsym,dsfont,array,layout,graphics,mathrsfs,braket,amsfonts,amsthm,amssymb,graphicx,subfigure,dcolumn,youngtab,color,bm,mathtools,braket,verbatim,url,cleveref,natbib,hypernat,extarrows,inputenc,multirow,soul}
\usepackage[mathlines]{lineno}
\usepackage{mathrsfs, xr-hyper}
\usepackage{physics}
\newcommand{\sg}[1]{}
\renewcommand{\sg}[1]{{\color{red}{#1}}} 
\newcommand{\pa}[1]{}
\renewcommand{\pa}[1]{{\color{blue}{#1}}} 
\begin{document}
\title{\Large Boosting lattice polarization \\ \normalsize Mixing the perspectives of geometry optimization and cell-augmentation}

\author{Pegah Azizi}
\affiliation{%
Department of Civil, Environmental, and Geo- Engineering,
University of Minnesota, Minneapolis, MN 55455, US}
\author{Rahul Dev Kundu}
\affiliation{%
Department of Civil and Environmental Engineering, University of Illinois at Urbana-Champaign, Urbana, IL, 61801, USA
}
\author{Xiaojia Shelly Zhang}
\email{zhangxs@illinois.edu}
\affiliation{%
Department of Civil and Environmental Engineering, University of Illinois at Urbana-Champaign, Urbana, IL, 61801, USA
}
\author{ Stefano Gonella}%
\email{sgonella@umn.edu}
\affiliation{%
Department of Civil, Environmental, and Geo- Engineering,
University of Minnesota, Minneapolis, MN 55455, US}%

\begin{abstract}

\textbf{Abstract:} 

Topologically polarized mechanical metamaterials enjoy a special built-in asymmetry that manifests as a preferred localization of edge states on selected edges. While this property has been shown for a few ideal Maxwell lattices, we currently lack systematic criteria to design families of structural systems exhibiting polarization. Here, we propose a framework to design polarized structural configurations enabled by topology optimization (TO), using both band and mode morphological properties as drivers of the optimization algorithm. Through the lens of TO, we are able to tap into a vast design space, unlocking geometric freedom far beyond what is achievable with canonical lattice architectures. At the same time, we elucidate important criteria that need to be satisfied, beyond the optimization outcome, to ensure robustness of the achieved polarization against perturbations of the edge morphology. These results provide the inspiration to loop back into the realm of ideal lattices in search of new configurations characterized by extreme polarization. The peculiar shape and connectivity of the TO-generated lattice offer a blueprint for identifying a new family of Maxwell trusses based on augmented kagome geometry. We demonstrate the achievement of strong polarization signatures up to a three-count edge state mismatch. For all the cases studied, we show agreement between theory, simulations, and experiments, which include laser vibrometry wave measurements on a waterjet-cut specimen and static tests on a 3D-printed prototype. \\

\vspace{-0.1in}
\end{abstract}

\maketitle

\section{Introduction}

Mechanical metamaterials are architected solids whose structure is engineered to achieve properties beyond those of conventional elastic media~\cite{CALLADINE_Pellegrino_mechanism,Lee_et_all_mech_mat_AdvMat_2012,Babaee_et_sll_metamaterial_2013,Rafsanjani_Pasini__metamaterials_2016,metamaterials_Bertoldi2017,Bolei_et_all_floppy_PNAS2020}. Recent advances at the intersection of topology and metamaterial design have led to the emergence of topological mechanical metamaterials, in which the mechanical response is governed by the topology of the bulk structure. This results in robustness against local imperfections, a feature commonly referred to as topological protection. Topological characterization can be carried out either in physical space, working on the real-space symmetry of the geometry, or in reciprocal $k$-space, through topological invariants of the system’s phonon band structure~\cite{Nash_etall_gyroscop_pnas_2015,He_acousTI_NatPhys_2016,Mousavi_etall_topoprotected_NatCom_2015,Yu_etall_topopseudo_NatCom_2018,Mou_etall_Valley_NatMat_2018,Vila_Pal_Ruzzene_PRB_VALLEY,Raj_Ruzzene_QVHE_NewJournalPhys_2017,Kane_Lubensky_Nphys_2014,Azizi-et-al_Fragile-kagome_PRL_2023,Azizi_2024_DW_PhysRevB}.

A class of systems of interest in this context is Maxwell lattices~\cite{Maxwell_1864,calladine_lattice_IJSS_1978,mao2018maxwell,jacobs1995generic}. In their ideal configuration, where nodes are connected by frictionless hinges, Maxwell lattices possess an equal number of constraints and degrees of freedom in the bulk (coordination number $z = 4$ in two dimensions), placing them on the verge of mechanical instability. According to the Maxwell--Calladine index, their linear response is governed by the balance between zero-energy modes (ZMs), which correspond to deformation mechanisms that cost no energy, and states of self-stress (SSSs), internal stress states that exert no net forces on the nodes~\cite{Zhang_2018fractopo,PaulosePNAStopo,Sun-et-al_kagome-lattices_PNAS_2012,lubensky2015phonons, Azizi-Gonella_Kagome-chain_PRApp_2024}. In topological Maxwell lattices, the ZMs can be directed to become exponentially localized at certain edges or interfaces, resulting in the ability to absorb localized loads without transferring significant stresses into the bulk~\cite{Kane_Lubensky_Nphys_2014,rocklin2017directional,ZhouPhysRevLettFloppy,Zunker-Gonella_Soft-lattice-wheel_EML_2021}. This phenomenon, known as topological polarization, grants the lattice an asymmetric mechanical response, where an excess of \textit{floppiness} on one edge is matched by a surplus of \textit{rigidity} on the opposite edge~\cite{rocklin2016mechanical,stenull2019signatures}. This polarization is captured by a polarization vector, whose orientation marks the direction along which floppy modes tend to exponentially localize. A key implication of the bulk--edge correspondence is that polarization is an intrinsic property of the bulk that manifests at the edges, and is thus independent of any trivial local morphological features of the edges. 

Translating the properties of polarization from the realm of ideal lattices to elastic metamaterials that can be realized via additive or subtractive manufacturing presents a significant challenge. Structural counterparts of Maxwell lattices inevitably involve finite-thickness elastic connections in lieu of ideal hinges, or require replacing trusses of rods (deformable only axially) with frames of beams that can deform in bending and shear. The finite stiffening and overconstraining introduced by these modifications have a diluting effect on the polarization predicted by ideal theory, causing weaker signatures and precluding rigorous mathematical modeling. Moreover, the kinematics of structural lattices cause the ZMs to migrate to finite frequencies, turning them into \textit{soft phonons} that often hybridize with bulk band modes~\cite{ma2018edge,charara2021topological,CHARARAPNAS2022}. As a result, the contrast between floppy and rigid boundaries reduces to an asymmetry between soft and stiff boundaries, whose weaker signatures nevertheless persist in the form of asymmetric wave transport.

A powerful approach to discover metamaterials with target properties is inverse design via topology optimization (TO)~\cite{Bendsoe1988,Bendsoe2004_TObook,Wang2021}. By distributing material within a design domain according to some performance criteria guided by optimization theory, TO allows finding families of complex configurations beyond intuition-driven architectures. TO has been successfully applied to design metamaterials with desired wave functionalities, including bandgap formation~\cite{Sigund-Jensen_TO-Bandgap_PRSA_2003, Dalklint2022_intro,Liu2024_intro, Dong2024_intro,Wang2019_ultrawideBG,Luo2022_soft,Wu2023_prescribedBG,Vatanabe2014_piezoBG,Jia2024_BG}, negative refraction~\cite{Christiansen2016_negativerefrac}, and, more recently, topological states~\cite{Christiansen2019_intro,Nanthakumar2019_intro,Chen2021_intro,Lu2021_intro,Christiansen2019_spinHall,Zhuang2022_intro,Zhang2022_intro,Luo2023_intro,AziziPnas2025}. In contrast, the design of polarized metamaterials has been primarily conducted through intuition-based methods and the design space has thus remained confined to canonical square~\cite{rocklin2016mechanical} and kagome lattices~\cite{rocklin2017transformable, Riva_et_all_topo_kagome_appliedphys_2018,mao2018maxwell,Sun-et-al_kagome-lattices_PNAS_2012} and, recently, augmented kagome configurations obtained via mirror-folding of classical kagome cells~\cite{chararaPRB2023AUGMENT}. The main obstacle against broader adoption of TO for this task is the difficulty to distill and effectively express mathematically the phenomenological attributes that govern polarization.

In this work, we provide a much-needed layer of generalization to the design of polarized metamaterials through a dual treatment that addresses in parallel the ideal and structural lattice space, with the two problems being connected and mutually inspiring. First, we put forth a TO-enabled approach for designing polarized metamaterials, in order to tap into the expansive design space available via optimization. To this end, we identify selected band and mode morphological properties, such as the spectral isolation of edge-localized branches and their unique spatial decay profiles, as the key requirements for polarization, and we implement them directly as objectives and/or constraints of the TO algorithm (Fig.~\ref{fig:1}a). We further endow the process with an a posteriori protocol (i.e., a series of structural checks to be performed on the results of the optimization) to verify the robustness of the polarization against different boundary trims. This step triages configurations in which the achieved asymmetry between the edges is not intrinsic to the bulk and may instead originate trivially from accidental edge features. The search for rationales that explain mechanistically how the TO algorithm achieves polarization prompts us to connect the TO-generated cell to some interpretable ideal configurations that satisfy Maxwell conditions, for which polarization can be formally defined. This inductive process leads us to discover a whole family of non-intuitive augmented kagome geometries with tunable built-in asymmetries, including extreme polarization signatures up to a winding number of $\mathrm{WN} = -3$~\cite{ShengqunAPL2025WN}. We experimentally confirm the soft--stiff edge dichotomy via laser vibrometry measurements on waterjet-cut prototypes (sec.~\ref{sec:laser}) and static tests on tabletop models (sec.~\ref{sec:lego}). 

\section{Topology optimization framework for lattice polarization}\label{sec:TO_framework}
\begin{figure*}[t]
\includegraphics[width=1.0\textwidth]{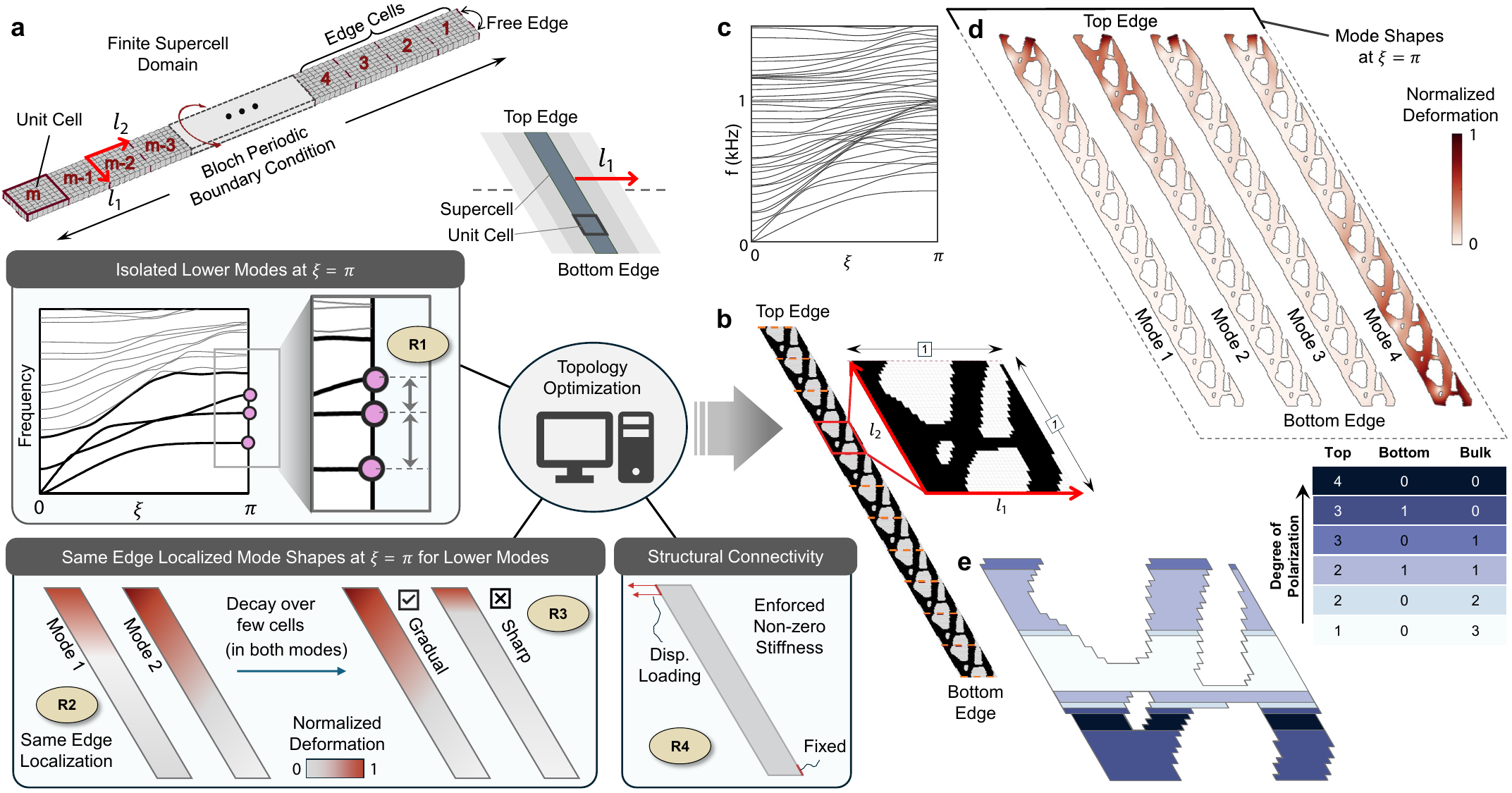}
\caption{\label{fig:1}Topology optimization framework to promote polarization of structural lattices. (a) Schematic of the design domain, comprising a finite supercell of $m$ unit cells, and pictorial rendition of the \textbf{R1-4} optimization requirements detailed in the main text. (b) Optimized supercell geometry with inset showing the constituent unit cell. (c) Computed band diagram of the optimized supercell structure. (d) Mode shapes for the first four eigenmodes at $\xi = \pi$, demonstrating consistent spatial localization at the top edge for the first three modes. The colorbar indicates normalized displacement magnitude. (e) Evolution of the degree of polarization for varying top-edge truncation levels, keeping the bottom edge fixed. The table lists the number of modes localized at the top/bottom edge or leaking into the bulk, at $\xi = \pi$, for different edge trims. The color scale (also reflected in the color coding of the edge trims) denotes the degree of polarization.}
\end{figure*}

In this section, we introduce the band and mode morphological properties selected as the key requirements for polarization. To explain this selection, we first recall a few rudiments of supercell Bloch analysis classically used to describe polarization, as well as some elements of topology optimization (TO) for Bloch wave problems. As illustrated in Fig.~\ref{fig:1}a, our design domain $\Omega$ is a supercell of finite length along primitive vector $\bm{l}_2$, consisting of $m$ (here, $m=11$) unit cells with identical geometry. In optimization, the cell repetition requirement is enforced by mapping the geometry of the first unit cell (as it emerges and is refined at every step of the process) to the rest of the cells. The supercell is then assumed to be repeated periodically along primitive vector $\bm{l}_1$ to capture the scenario of an infinite strip parallel to $\bm{l}_1$ and finite along $\bm{l}_2$. The dynamical analysis boils down to the one-dimensional (1D) problem of waves traveling with scalar wavenumber $\xi$ along the strip. Accordingly, we enforce Bloch-periodic boundary conditions along $\bm{l}_1$ and free boundary conditions in the transversal direction. To describe the geometry of the supercell, we use a scalar field $\boldsymbol{z} \in [0,1]$ that assigns solid or void elements within the design domain with values $z_e=1$ or $z_e=0$, respectively. For a given supercell geometry parameterized by $\boldsymbol{z}$, we can construct the stiffness and mass matrices $\mathbf{K}(\boldsymbol{z})$ and $\mathbf{M}(\boldsymbol{z})$, and subsequently obtain the eigenpair $\{\lambda^{(i)}_\xi(\boldsymbol{z}), \boldsymbol{\phi}^{(i)}_\xi(\boldsymbol{z})\}$ for the \textit{i}-th mode at wavenumber $\xi$ %by solving the eigenvalue problem 
that satisfy the systems of equations $\left(\mathbf{\hat{K}}_\xi(\boldsymbol{z}) - \lambda^{(i)}_\xi\mathbf{\hat{M}}_\xi(\boldsymbol{z})\right)\bm{\phi}^{(i)}_\xi = \bm{0}$, where $\mathbf{\hat{K}}_\xi$ and $\mathbf{\hat{M}}_\xi$ respectively denote the reduced forms of the stiffness and mass matrices considering Bloch periodic boundary conditions evaluated at wavenumber $\xi$, $\lambda^{(i)}_\xi$ is the square of the $i$-th modal frequency, and $\bm{\phi}^{(i)}_\xi$ is the $i$-th mode shape. We also assume the eigenvectors $\boldsymbol{\phi}^{(i)}_\xi$ are normalized w.r.t. $\mathbf{\hat{M}}_\xi$, i.e., $\boldsymbol{\phi}^{(i)}_\xi\cdot\mathbf{\hat{M}}_\xi\boldsymbol{\phi}^{(i)}_\xi=1$. Next we define the following functions based on the eigenpairs $\{\lambda^{(i)}_\xi(\boldsymbol{z}), \boldsymbol{\phi}^{(i)}_\xi(\boldsymbol{z})\}$:
\begin{equation}\label{eq:modal_functions}
\setlength{\jot}{5pt}
\begin{aligned}
& \tau^{(i)}_\xi\left(\lambda^{(i)}_\xi(\boldsymbol{z}), \lambda^{(i+1)}_\xi(\boldsymbol{z})\right)=\lambda^{(i+1)}_\xi-\lambda^{(i)}_\xi, \\
& \eta^{(i,k)}_\xi\left(\boldsymbol{\phi}^{(i)}_\xi(\boldsymbol{z})\right) = \dfrac{\sum_{j=1}^{k}h_\xi^{(i,j)}(\boldsymbol{\phi}^{(i)}_\xi)}{\sum_{j=1}^{m}h_\xi^{(i,j)}(\boldsymbol{\phi}^{(i)}_\xi)}, \\
& \chi^{(i,k)}_\xi\left(\boldsymbol{\phi}^{(i)}_\xi(\boldsymbol{z})\right) = \dfrac{h_\xi^{(i,k)}(\boldsymbol{\phi}^{(i)}_\xi)}{\sum_{j=1}^{k}h_\xi^{(i,j)}(\boldsymbol{\phi}^{(i)}_\xi)}, \\
\end{aligned}
\end{equation}
where, $\tau^{(i)}_\xi$ denotes the difference of the eigenvalues between the $i$-th and $(i+1)$-th modes at wavenumber $\xi$; $\eta^{(i,k)}_\xi$ is the 1-k cell localization index, which captures the extent of deformation in the 1-k cells relative to the whole supercell for mode shape $\boldsymbol{\phi}^{(i)}_\xi$; $\chi^{(i,k)}_\xi$ is the k-th cell deformation participation factor, which quantifies the relative contribution of the k-th cell in the total deformation of cells 1-k for mode shape $\boldsymbol{\phi}^{(i)}_\xi$. The quantity $h_\xi^{(i,j)}=\boldsymbol{\phi}^{(i)}_\xi\cdot\mathbf{D}^{(j)}\boldsymbol{\phi}^{(i)}_\xi$ is a deformation index defined as the sum of the squares of the $i$-th mode shape $\boldsymbol{\phi}^{(i)}_\xi$ components for all degrees of freedom (DOFs) in the $j$-th periodic unit, computed using a diagonal matrix $\mathbf{D}^{(j)}$ where the diagonal entries are 1 if the corresponding DOFs belong to the $j$-th periodic unit. 

 %made minor edits for Fig 1b

To obtain a polarized supercell, we target the following requirements simultaneously: \\
\noindent (\textbf{R1}) We want to isolate the lowest two bands from the higher bands at $\xi=\pi$. This is motivated by the observation that, while structural lattices do not support zero modes per se, the structural counterparts of polarized lattices do feature soft finite-frequency modes whose branches are spectrally located well below the bulk modes. As these modes flatten significantly as we approach $\xi=\pi$, we observe a gap between these branches and the so-called bulk band in the short-wavelength limit. \\
\noindent (\textbf{R2}) We want to ensure that the modal energies for the first two mode shapes are concentrated at the same edge. This enforces simultaneously the localization of deformation at the edges typical of floppy edge modes and the excess of softness on one edge required by polarization. \\ 
\noindent (\textbf{R3}) Enforcing only requirements \textbf{R1} and \textbf{R2} often promotes the emergence of quasi-dangling features of the edges and trivial soft mechanisms (see Fig.~\ref{fig:S1_a} in SI~\ref{addl_TO_designs}). This occurrence is reflected by excessive flattening of the lower bands and over-concentration of modal energies at the soft edge. Therefore, we prevent these trivial soft edge modes by restricting the allowed isolation of the lower bands and by promoting a gradual decrease of modal energies within the first few units. \\
\noindent (\textbf{R4}) In addition, we consider a separate quasi-static loading scenario for the supercell to enforce a minimum structural stiffness and guarantee preservation of lattice connectivity throughout the optimization process.

These requirements are consolidated into an optimization formulation as: 
\begin{equation}\label{eq:TO_problem}
\setlength{\jot}{5pt}
\begin{aligned}
\min\limits_{\bm{z}} \quad
& -\sum\limits_{p=1}^{2}\alpha_\tau^{(p)}\tau^{(p)}_\pi(\bm{z}) + \alpha_v v(\bm{z}), \\ 
\quad \ \ \text { s.t. } \quad
& \eta_{\pi}^{(p,k)}(\bm{z}) - \varepsilon_\eta \geq 0, \quad p=1,2, \quad k\in\mathcal{S}_{2-m}, \\ 
& \chi_{\pi}^{(p,k)}(\bm{z}) - \varepsilon_\chi \geq 0, \quad p=1,2, \quad k\in\mathcal{S}_{2-m}, \\ 
& \tau^{(p)}_\pi (\bm{z})/\tau^{*}(\bm{z}) - \varepsilon_\tau \leq 0, \quad p=1,2, \\ 
& c(\bm{z})/c^{*} - \varepsilon_c \geq 0, \\ 
& v(\bm{z}) - \varepsilon_v \leq 0, \\ 
& 0 \leq z_e \leq 1, \quad e=1, \ldots, N_e,\\ 
\quad \ \ \text { with } \quad
& \left(\mathbf{\hat{K}}_\pi(\bm{z})) - \lambda^{(i)}_\pi\mathbf{\hat{M}}_\pi(\bm{z}))\right)\bm{\phi}^{(i)}_\pi = \bm{0}, \quad i=1,2,3 \\
& \mathbf{{K}}(\bm{z}) \mathbf{u} = \bm{0}, \quad \text{ with } \mathbf{u}=\mathbf{u}_0 \text{ on } \partial\Omega,\\
\end{aligned}
\end{equation}
where, the quantities $\eta, \chi$ and $\tau$ are obtained from the Bloch eigenvalue problem (second-to-last line in Eq.~\eqref{eq:TO_problem}) as per the definitions in Eq.~\eqref{eq:modal_functions}, $c({\mathbf{u}}(\bm{z}))=\sum_{e} \bm{u}_e \cdot \bm{k}_e \bm{u}_e$ is the corresponding compliance for the static problem (last line in Eq.~\eqref{eq:TO_problem}),  $\mathbf{u}$ is the calculated displacement field with prescribed displacement $\mathbf{u}_0$ on the domain boundary $\partial \Omega$, $v(\bm{z})=\sum_{e}z_e v^0_e/\sum_{e}v^0_e$ is the solid volume fraction with $v^0_e$ being the volume of solid element $e$, $\alpha_\tau$ and $\alpha_v$ are the objective component weights, $\mathcal{S}_{2-m}$ is the set of integers between 2 and $m$, $\tau^{*}$ and $c^{*}$ are respectively the normalization factors for eigenvalues and compliance, and $\varepsilon_\eta, \varepsilon_\chi, \varepsilon_\tau, \varepsilon_c$, and $\varepsilon_v$ are the constraint tolerances.  

In the optimization formulation~\eqref{eq:TO_problem}, the requirement \textbf{R1} is achieved through the objective function that maximizes the gaps between branches 1 and 2 and branches 2 and 3 at $\xi=\pi$ for a spectral isolation of these modes from the bulk bands. The requirement \textbf{R2} is realized using the first constraint set on $\eta$. By confining the deformation of the first two modes within the first k cells, this constraint set simultaneously promotes localization and enforces that such localization appears at the same edge for both modes. Next, in order to fulfill the requirement \textbf{R3}, we impose the second constraint set on $\chi$ and the third constraint set on $\tau$, which ensure sufficient modal deformation in the k-th cell and prevent large isolation of the first two bands, respectively. Together, these two constraint sets prevent overly localized modal energies within the first cell - a scenario often trivially associated with soft local features of the edges. The fourth constraint on $c$, in conjunction with a static displacement-controlled problem (last line of Eq.~\eqref{eq:TO_problem}), ensures a sufficient level of compliance to maintain supercell connectivity, thereby fulfilling the requirement \textbf{R4}. In addition, we use volume penalization in the objective function and add a fifth constraint on $v$ to promote a lightweight design. Note that, the presented optimization formulation~\eqref{eq:TO_problem} is a general one, and can be tailored to specific design needs by setting different values for $\alpha_\tau,\,\alpha_v,\,k,\,\varepsilon_\eta,\,\varepsilon_\chi,\,\varepsilon_\tau,\,\varepsilon_c,\,\varepsilon_v,\,\tau^{*},$ and $c^{*}$. Here, we use $\alpha_\tau=0.5,\,\alpha_v=0.001,\,k=4,\,\varepsilon_\eta=0.8,\,\varepsilon_\chi=0.045,\,\varepsilon_\tau=4,\,\varepsilon_c=0.8,\,\varepsilon_v=0.6,\,\tau^{*}=\tau_\pi^{(3)}$, and $c^{*}$ equals the compliance value at first optimization iteration. Most of these parameter choices, especially the ones depending on the eigenvalues and eigenvectors, are inspired by the behavior of a structural kagome design (see Fig.~\ref{fig:S1_b} in SI~\ref{addl_TO_designs}). The optimization problem is solved using a gradient-based optimizer, the Method of Moving Asymptotes~\cite{Svanberg1987_IJNME}, and the related gradients for eigenvalues and eigenvectors are computed following the Nelson method~\cite{Nelson1976_AIAA,Lee1996_JSV,Xue2022_CMAME}. 

 \begin{figure*}[t]
\includegraphics[width=1.0\textwidth]{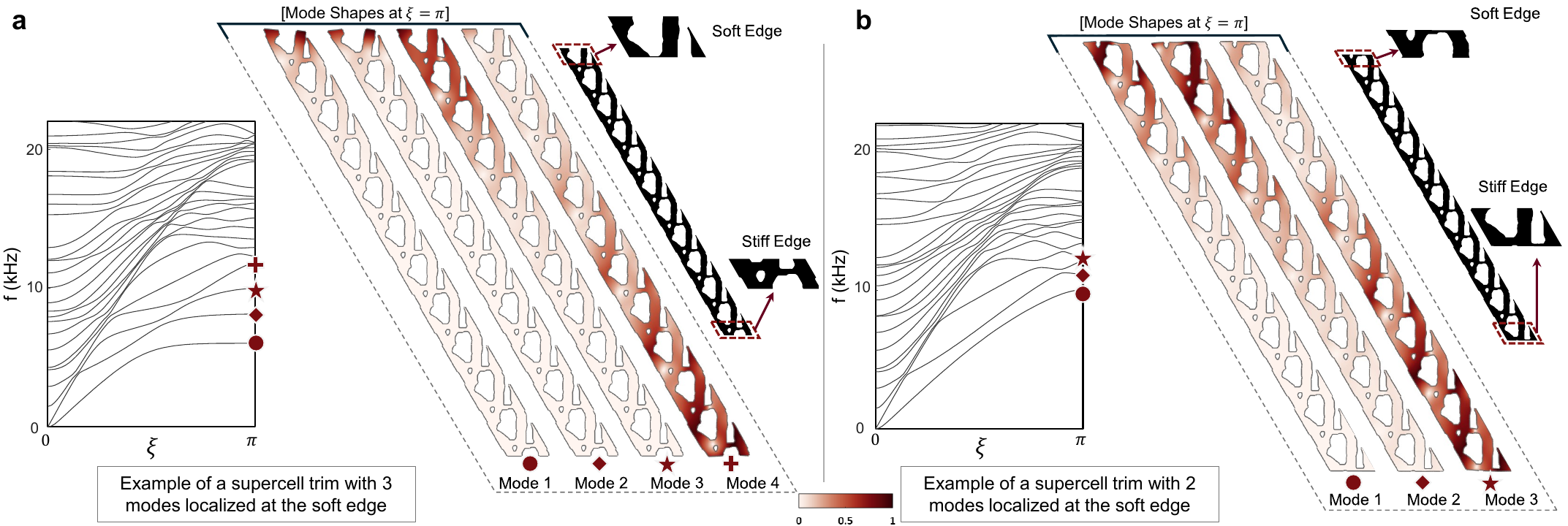}
\caption{\label{fig:2}Complete supercell analysis of the optimized designs (refined and smoothened). (a) Band diagram and first four mode shapes (evaluated at $\xi = \pi$) for a strongly polarized supercell edge trim that exhibits three modes localized at the top edge. Insets provide zoomed views of the \textit{soft} (top) and \textit{stiff} (bottom) edge geometries. (b) Comparative analysis for an alternative trim, demonstrating that localization at the soft edge persists for the first two modes even with a trim that removes all protruding edge features. In both panels, red markers identify spectral points in the dispersion plots at which the mode shapes are computed. The colorbar represents normalized displacement magnitude.}
\end{figure*}

\begin{figure*}[t]
\includegraphics[width=1.0\textwidth]{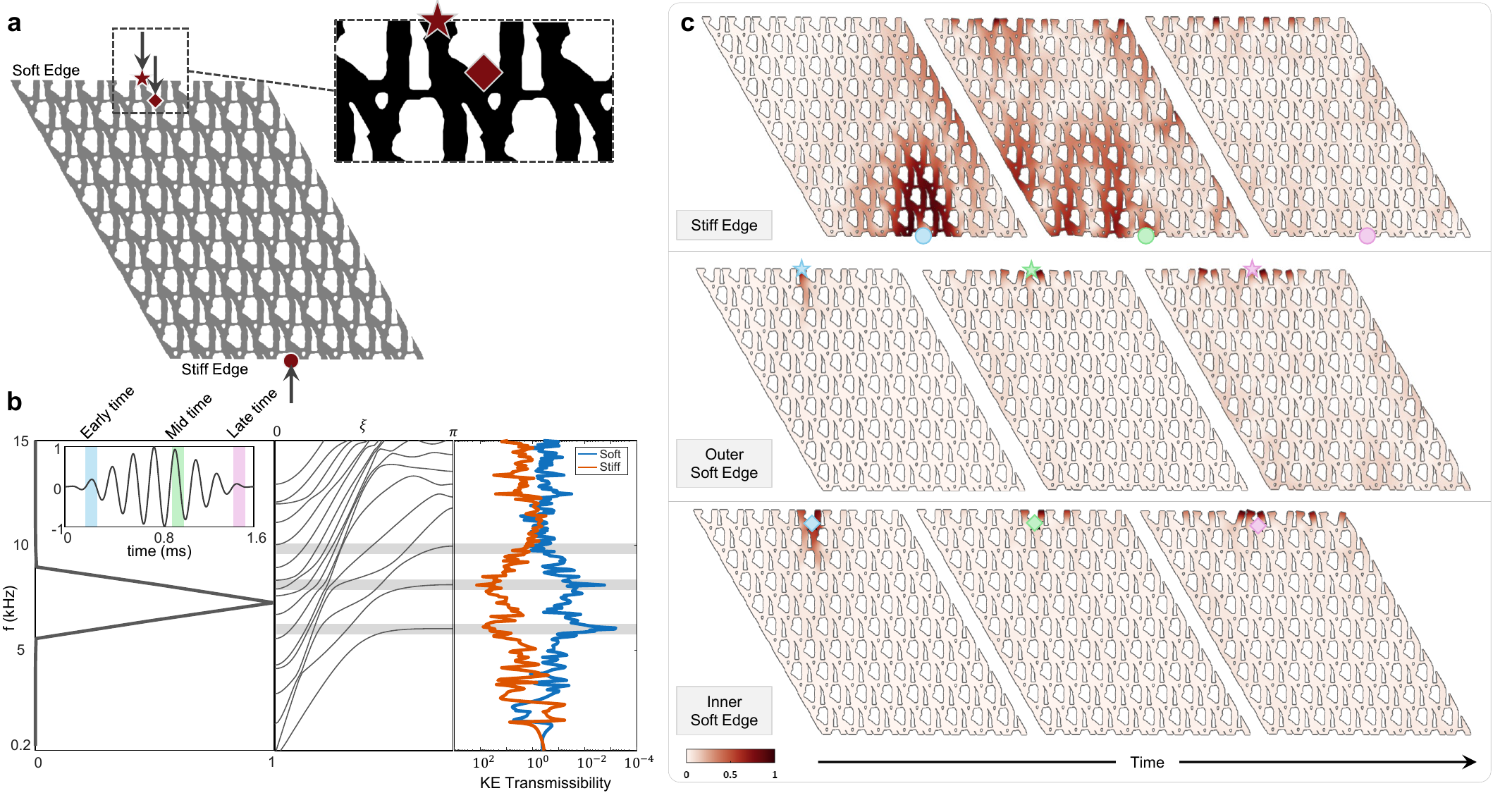}
\caption{\label{fig:3}Full-scale simulation and energy localization in the optimized polarized lattice. (a) 2D tessellation highlighting the \textit{soft} (top) and \textit{stiff} (bottom) edges. Red markers denote point-force excitation locations at the stiff edge (circle), outer soft edge (star), and inner soft edge (diamond). (b) (Left) 9-cycle tone burst excitation with a carrier frequency of 7.7~kHz and its frequency spectrum. (Center) Supercell band diagram with gray shaded bands highlighting frequencies of the first three modes at $\xi = \pi$. (Right) Computed kinetic energy (KE) transmissibility across the frequency range 0.2–15 kHz for excitations applied at the soft (blue) and stiff (orange) edges. (c) Transient wavefield snapshots at early (blue), mid (green), and late (pink) times, for excitation at the stiff edge, outer soft edge, and inner soft edge, respectively. The results demonstrate robust energy localization at the soft edge, regardless of whether excitation is applied at the outer or inner boundary. The colorbar denotes displacement magnitude normalized by the global maximum across all time steps for each excitation scenario.}
\end{figure*}

The supercell geometry generated by the algorithm and its constituent unit cell (normalized to unit side length) are presented in Fig.~\ref{fig:1}b with the  resulting band diagram in Fig.~\ref{fig:1}c. We focus our attention on the low-frequency region of the band diagram, where the bands appear to be more rarefied, and we focus on the four lowest branches. To assess whether any of these are edge-localized states, we examine their supercell mode shapes evaluated at $\xi = \pi$ (Fig.~\ref{fig:1}d): we find that the first three modes are edge-localized, whereas the deformation in mode 4 leaks into the bulk. Moreover, the localization observed in modes 1--3 consistently occurs at the top edge, indicating that a pronounced asymmetry in edge softness has indeed been achieved. At this stage, it is important to check whether the achieved asymmetry results solely from the geometry of the unit cell, and can therefore be seen as an intrinsic property of the bulk, or stems instead from specific shape features of the edges. In the former case, it can be seen as a true structural counterpart of the polarization observed in topological lattices, and can enjoy robustness against changes (by design or accidental) of the edge morphology. Otherwise, it can be ascribed to trivial edge morphological effects, and, as such, it does not enjoy any protection, being diluted or vanishing altogether when the edges are modified. 

To assess the configurational robustness of this asymmetric response, we perform the following sensitivity analysis. Starting from the TO-generated configuration, we generate a family of supercells by progressively trimming the top boundary and removing successive layers of elements. This operation eliminates protruding edge features whose local dynamics may dominate the response and produce trivial localization effects, while leaving the bottom edge unchanged. For each configuration, we inspect the supercell mode shapes of modes 1–4 and classify them as top-edge-localized, bottom-edge-localized, or bulk modes. For instance, a 2-1-1 configuration has two modes localized at the top edge, one at the bottom edge, and one bulk mode. The results are listed in the table in Fig.~\ref{fig:1}e and ranked (and color-coded accordingly from dark to light blue) based on the strength of polarization. We classify the strength of polarization in terms of two factors: (i) number of bulk modes in the set, and (ii) imbalance between top- and bottom-edge localized modes. The former metric is especially important if we think of the practical manifestation of polarization: in a polarized lattice, we expect the structure to act as a (mechanical) conductor when excited from the stiff edge and as an insulator from the soft edge, where deformation localizes. For this to work, it is necessary that no bulk mode can be activated by an excitation applied on the soft edge, since such mode would immediately result in some leakage into the bulk. Our classification treats the number of bulk modes (from 0 for the most polarized case to 3 for the least polarized) as primary parameter and the imbalance between top and bottom as secondary parameter. Based on this criterion, for example, the \textit{fully polarized} 4-0-0 configuration with 4 modes localized at the top edge, no bulk modes (and pronounced edge imbalance) is ranked highest, followed by the 3-1-0 case that displays a weaker imbalance, the 3-0-1 case characterized by some bulk leakage, etc. Moreover, the polarization of the 3-0-1 case is deemed stronger than that of the 2-1-1 case, since they both have one bulk mode but different edge imbalance levels. Notably, in the 3-1-0 (2-1-1) configuration, the bottom-edge localized mode appears at a higher frequency than the top-edge localized modes; i.e., the first three (two) lowest-frequency modes remain localized at the top edge. In summary, different trims result in different strengths of polarization, and only a small subset guarantees full polarization.

A few considerations are in order. While TO succeeds at achieving edge localization imbalance and separating localizing modes from the bulk band, TO alone cannot guarantee that the polarization is truly a bulk property, and it cannot quantify the strength of polarization. To ensure these additional attributes, the optimization step must be followed by an \textit{a posteriori} evaluation that considers all possible supercell trims and ranks them according to strength of polarization, eliminating deficient designs. In the case shown, this analysis reveals that, across all trims, the top edge remains consistently more compliant than the bottom one, and preference for localization at the top edge is never entirely suppressed. Even in the least polarized scenarios, the fundamental mode remains localized at the top edge. In essence, this configuration is characterized by strong edge asymmetry, minimal risks of bulk leakage, and robustness of these effects against boundary perturbations. It is important to note that, in other designs, the outcome can be different. Different trimming of the TO-generated supercell would cause dilution of the polarization in the form of bulk leakage, re-balancing of the edges or even inversion of the asymmetry (more localization at the bottom edge). Examples of such cases are given in the SI~\ref{addl_TO_designs}.  

\section{Computational and experimental validation of the optimized design\label{sec:laser}}

\begin{figure*}[t]
\includegraphics[width=1.0\textwidth]{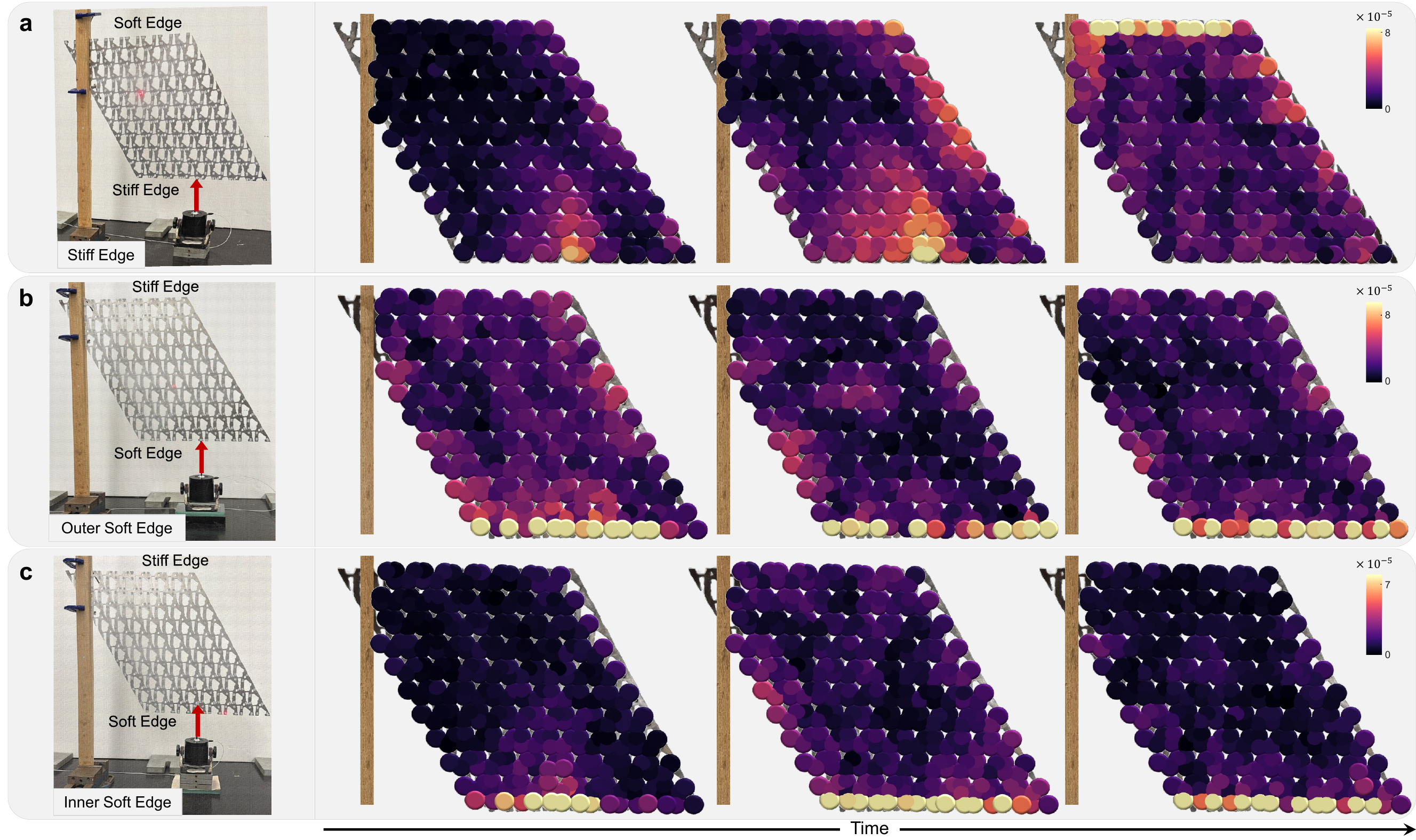}
\caption{\label{fig:4}Snapshots of wavefields experimentally acquired via Scanning Laser Doppler Vibrombeter (SLDV). Transient energy transport for excitations at the (a) stiff edge, (b) outer soft edge, and (c) inner soft edge, for a carrier frequency of 7.7~kHz. The color scale represents the instantaneous amplitude envelope of the in-plane velocity field, normalized to the $99.95^{\mathrm{th}}$ percentile of the maximum global envelope to suppress transient noise spikes. In each row, the leftmost panel illustrates the specimen orientation and boundary conditions, with the lattice clamped at the left boundary. Red arrows denote the location and orientation of the shaker stinger.}
\end{figure*}

We now proceed to characterize the finite-frequency manifestation of the polarization exhibited by the TO-generated configuration. We post-process the jagged boundaries produced by TO to obtain a smooth geometry and scale the unit-cell dimensions so that each unit-cell side measures 5~cm, thereby achieving a lattice size compatible with manufacturing constraints (see SI~\ref{laserexperiment} for additional details regarding the refined geometry). Fig.~\ref{fig:2}a depicts an 11-cell supercell with the \textit{soft} (top) and \textit{stiff} (bottom) boundaries highlighted. The computed band diagram and the corresponding mode shapes at $\xi = \pi$ confirm the persistence of the predicted edge asymmetry, with the first three modes clearly localized at the soft boundary. To confirm that this polarization is an intrinsic bulk property rather than a trivial artifact of the edge geometry, we consider an alternative supercell trim in which all the protruding edge elements are removed (Fig.~\ref{fig:2}b). Despite this modification, the soft--stiff edge imbalance is overall preserved (neither erased nor reversed), as deformation remains localized at the top edge for the 
lowest modes. On the other hand, we recognize that the two trims exhibit different levels of polarization strength and some dilution of the asymmetry. 

To appreciate the manifestation of asymmetry at the finite lattice scale, we perform transient wave propagation simulations on a $10 \times 8$ unit-cell domain (Fig.~\ref{fig:3}a). The system is excited by an in-plane nine-cycle tone burst with a carrier frequency of 7.7~kHz (Fig.~\ref{fig:3}b, left), chosen to coincide with the spectral range of the localized edge modes identified in the supercell band diagram (Fig.~\ref{fig:3}b, center). We consider excitations applied at the stiff edge (circle) and at the outer soft edge (star marker). The results at three stages of propagation (blue, green, and pink) reveal a profound dichotomy in the dynamic response. Specifically, waves launched from the stiff edge readily propagate through the bulk 
(Fig.~\ref{fig:3}c, first row); notably, at later times, the energy eventually localizes at the opposite soft boundary, where it is ultimately trapped. Conversely, waves fired at the soft boundary remain confined to the edge throughout the simulation, with minimal penetration into the bulk (Fig.~\ref{fig:3}c, second row). To further verify that the observed localization is not trivially induced by deformation of protruding beam-like edge elements that may behave as local resonators, we perform an additional simulation in which the excitation is applied at the inner soft edge (diamond marker), located half a unit cell into the lattice interior. 
We see that the response remains strongly localized near the soft edge (Fig.~\ref{fig:3}c, third row). See also Supplementary Video 1 illustrating the three scenarios. We also evaluate the kinetic energy (KE) transmissibility over the 0.2–15 kHz frequency range (Fig.~\ref{fig:3}b, right) for excitations applied at the soft (blue) and stiff (orange) edges. Shaded bands mark the frequencies of the first three modes at $\xi = \pi$, whose mode shapes display top-edge localization. The mismatch between the curves graphically captures the degree of polarization~\cite{CHARARAPNAS2022}; notably, the strongest signature occurs near the flat region of the first two modes, for $\xi \longrightarrow \pi$.

We now seek experimental validation of the numerically predicted asymmetric wave transport between the stiff and soft boundaries. To this end, we perform 3D scanning laser Doppler vibrometry (SLDV) measurements on a waterjet-cut specimen (see SI~\ref{laserexperiment} for fabrication details, experimental setup, and post-processing procedures). The experiments directly replicate the numerical excitation scenarios, employing the same nine-cycle tone-burst signal with a carrier frequency of 7.7~kHz for excitations applied at the stiff edge (Fig.~\ref{fig:4}a), outer soft edge (Fig.~\ref{fig:4}b), and inner soft edge (Fig.~\ref{fig:4}c). The specimen orientation, clamping, and stinger polarization are illustrated in the left panels of Fig.~\ref{fig:4}, for each test. To visualize transient energy transport, we reconstruct the instantaneous amplitude envelope from the measured in-plane velocity fields using Hilbert-transform post-processing (see SI~\ref{laserexperiment} for details). Representative wavefield snapshots over time for each excitation case are presented in Fig.~\ref{fig:4}. The color scale is normalized by the $99.95^{\mathrm{th}}$ percentile of the global maximum envelope across all spatiotemporal measurements of each test, thereby suppressing isolated noise spikes while preserving the dynamic range of the propagating field. The experimental results reveal a pronounced dichotomy in wave transport, in strong agreement with the simulations. Excitation from the stiff edge generates a bulk wave that traverses the lattice and subsequently accumulates at the opposite soft edge. In contrast, excitation from either soft-edge location produces strongly localized motion confined to the boundary, effectively shielding the lattice interior (see also Supplementary Video 2). 

\section{Topological polarization of TO-inspired Maxwell lattices\label{sec:lego}}

While the power of TO is precisely to generate configurations of arbitrary complexity beyond the bounds of known canonical geometries, it is interesting to seek a mechanistic rationale for why a given configuration is ultimately found, i.e., understand what are the key geometric properties that satisfy the imposed design requirements. Inspired by this goal, we inspect the tessellated structure shown in Fig.~\ref{fig:3}a. The lattice can be interpreted as a modified structural kagome lattice with augmented connectivity: the unit cell resembles an asymmetric twisted kagome cell composed of two scalene triangles connected by thin structural ligaments, accompanied by two additional slender connectors (one thin and one thick) that bridge the kagome cell to its neighbors. Alternatively, the lattice can be viewed as a tessellation of kagome cells interspersed with irregular pentagonal voids. In this sense, we recognize, at least qualitatively, the structural counterparts of two geometric ingredients observed in topologically polarized Maxwell lattices: (i) the low-symmetry bi-triangular cell of topological kagome lattices, and (ii) a macro-cell generated by \textit{augmenting} a kagome cell, reminiscent of the strategies employed for bi- and quad-kagome lattices in~\cite{chararaPRB2023AUGMENT}.

\begin{figure*}[t]
\includegraphics[width=1.0\textwidth]{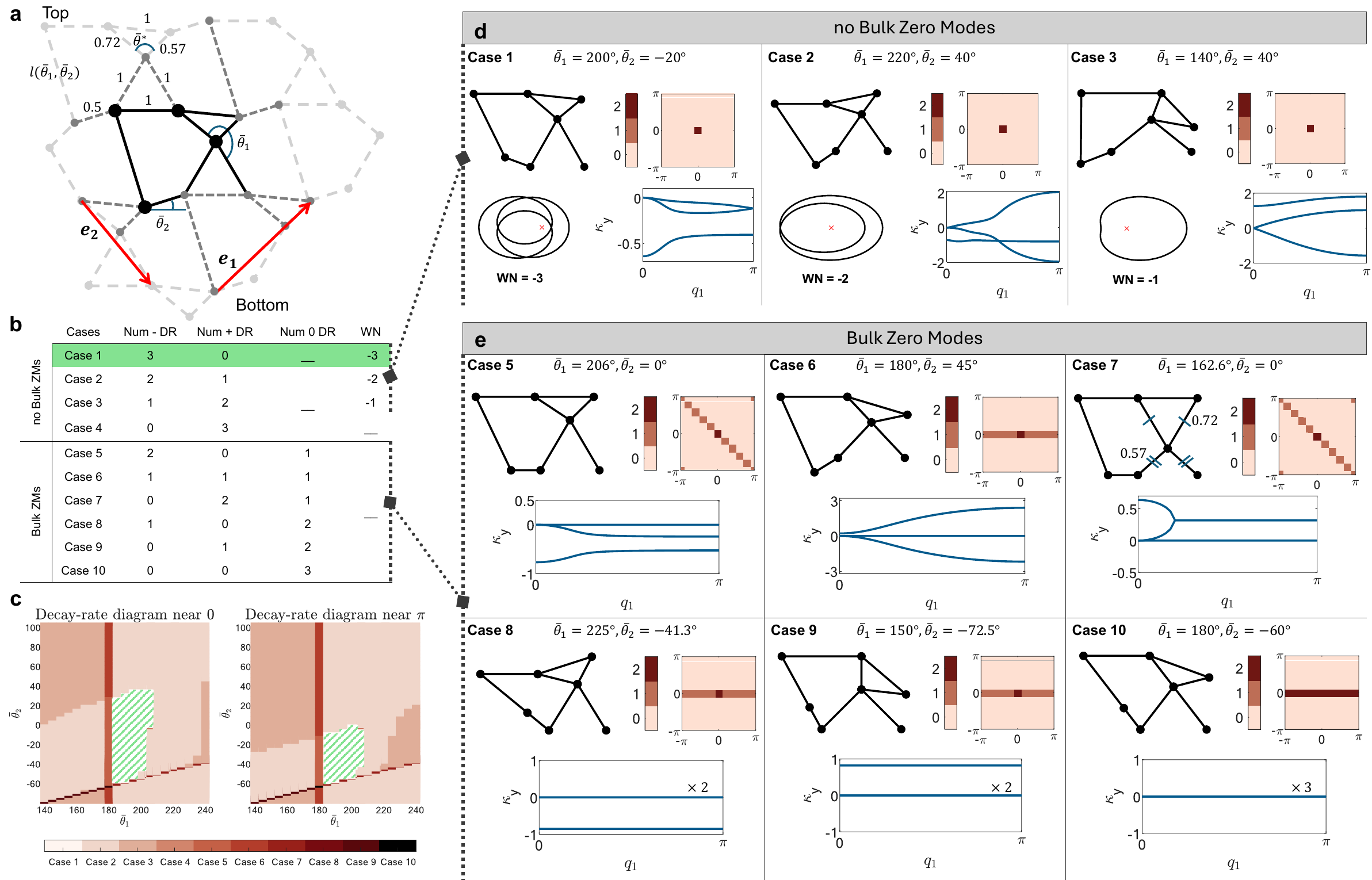}
\caption{\label{fig:5}Zero-mode analysis and topological phase landscape of the augmented kagome lattice inspired by TO-geneated configuration. (a) Unit cell with primitive vectors $\mathbf{e}_1$ and $\mathbf{e}_2$ and configurational parameters $\bar\theta_1$ and $\bar\theta_2$. Nodes and bonds belonging to the primary unit cell are denoted by black circles and solid lines, respectively, while dashed gray lines indicate connections to adjacent cells. (b) Classification of ten distinct Cases based on the number of positive, negative, and zero decay rate (DR) modes, and the corresponding winding number (WN) (when defined). 
Cases 1--4 are characterized by the absence of bulk ZMs, whereas Cases 5--10 exhibit at least one bulk ZM. (c) Topological phase diagrams depicting ZM decay-rate distributions as a function of the angular parameters ($\bar\theta_1, \bar\theta_2$), evaluated near $q_1=0$ and $q_1=\pi$. Case 1 (shaded green) represents the fully polarized lattice phase, featuring three negative DRs and $\mathrm{WN} = -3$. (d--e) Selected representative configurations: (d) without bulk ZMs and (e) with bulk ZMs. For each Case, subplots present the unit-cell geometry, the nullity of the compatibility matrix across the Brillouin zone, the WN $(\nu(q_1))$, and the ZM decay rate $(\kappa_y)$ (along the $\mathbf{e}_2$-direction) as a function of the wavenumber, $q_1$, (along the $\mathbf{e}_1$-direction). The presence of bulk ZMs (e) is associated with collinear structural mechanisms, which mark the topological phase transition boundaries in (c)).}
\end{figure*}

These observations motivate the construction of a new Maxwell lattice inspired directly by the geometric characteristics of the TO-generated cell, with the goal of investigating the conditions under which the lattice becomes polarized. Working with an ideal lattice allows a rigorous treatment of polarization, including the computation of topological invariants. To this end, we assemble a unit cell consisting of a truss of rods, arranged according to the edges of the triangles and struts identified in the TO configuration, connected by frictionless hinges. A general realization of this unit cell is depicted in Fig.~\ref{fig:5}a, where solid black lines denote primary cell bonds and dashed gray lines represent connections to adjacent unit cells. In an infinite periodic medium, each node maintains four-fold connectivity, satisfying the Maxwell criterion in two dimensions (average coordination number $z = 4$). The geometry is defined by an equilateral triangle of unit side length and a scalene triangle with side lengths (1, 0.57, 0.72), rotated relative to one another by the angle $\bar\theta_1$ and adjoined to an irregular pentagon. The pentagonal shape features a base side of fixed length 0.5 inclined at an angle $\bar\theta_2$. Under these geometric constraints, the length of the long slanted segment $l$ is uniquely determined by the pair of angular parameters $\bar\theta_1$ and $\bar\theta_2$. The overall configuration is therefore fully parameterized by these two angular variables. The corresponding primitive lattice vectors (indicated in Fig.~\ref{fig:5}a) are given by: 

\vspace{-0.1in}
\[
\mathbf{e}_1=\frac{1}{2}
\left(
\begin{array}{c}
\cos\bar\theta_2 + 1 + 1.14\,\cos(\bar\theta_1-60^\circ-\bar\theta^*)\\
\sin\bar\theta_2 + \sqrt{3} + 1.14\,\sin(\bar\theta_1-60^\circ-\bar\theta^*)
\end{array}
\right),
\]
\vspace{-0.15in}
\[
\mathbf{e}_2=-\frac{1}{2}
\left(
\begin{array}{c}
1.44\,\cos(\bar\theta_1-60^\circ)-1\\
1.44\,\sin(\bar\theta_1-60^\circ)+\sqrt{3}
\end{array}
\right),
\]
where $\bar\theta^*=101^\circ$ is the interior angle opposite the unit-length side in the scalene triangle. 

To investigate the asymmetric response of this augmented configuration, we recall and briefly summarize the treatment of topological polarization in Maxwell lattices~\cite{mao2018maxwell}. For a frame consisting of $N$ nodes connected by $N_c$ bonds in $d$ dimensions, the compatibility matrix $\mathbf{C}$ relates the $Nd$-dimensional vector of nodal displacements $\mathbf{U}$ to the $N_c$-dimensional vector of bond extensions $\mathbf{E}$, i.e., $\mathbf{C}\mathbf{U}=\mathbf{E}$. For a bond connecting nodes $n$ and $m$, the extension is given by $E_{nm} = \mathbf{a}_{nm}\cdot(\mathbf{U}_n - \mathbf{U}_m)$, where $\mathbf{a}_{nm}$ is the unit vector from node $m$ to $n$. The null space of $\mathbf{C}$ ($\mathbf{C}\mathbf{U}=\mathbf{0}$) identifies the set of ZMs that produce no change in bond length. The number of ZMs $N_0$ is determined by the rank--nullity relation $N_0 = Nd - \text{rank}(\mathbf{C})$~\cite{mao2018maxwell}. 
In periodic lattices, $\mathbf{C}$ is block-diagonalized at each wave vector $\mathbf{q} = (q_1, q_2)$. In our problem, the unit cell contains four nodes (black circles in Fig.~\ref{fig:5}a) with eight degrees of freedom and eight bonds (three internal and five intercellular), yielding a square $8 \times 8$ matrix $\mathbf{C}(\mathbf{q})$. The existence of ZMs is verified by enforcing the condition $\det \mathbf{C}(q_1, q_2) = 0$. To assess whether a ZM is a bulk mode or an edge-localized mode, we consider a wave propagating along one lattice direction (e.g., denoted by the wavenumber $q_1$) and examine whether the mode experiences exponential decay along the transverse direction $q_2$. Specifically, we consider a strip geometry that is infinite along the propagation direction and finite, with open boundaries, along the transverse direction. In this setting, exponential decay along $q_2$ indicates localization at one of the strip boundaries, with the sign of the decay determining the localized edge. For a given $q_1$, the roots of $\det \mathbf{C}(q_1,q_2)=0$ yield generally complex-valued wavenumbers $q_2$, whose imaginary component, $\kappa_y=\mathrm{Im}(q_2)$, defines the spatial decay rate (DR) of the mode into the bulk. Modes with $\kappa_y=0$ correspond to bulk ZMs, whereas modes with $\kappa_y\neq0$ are exponentially localized at the edge.

\begin{figure*}[t]
\includegraphics[width=1.0\textwidth]{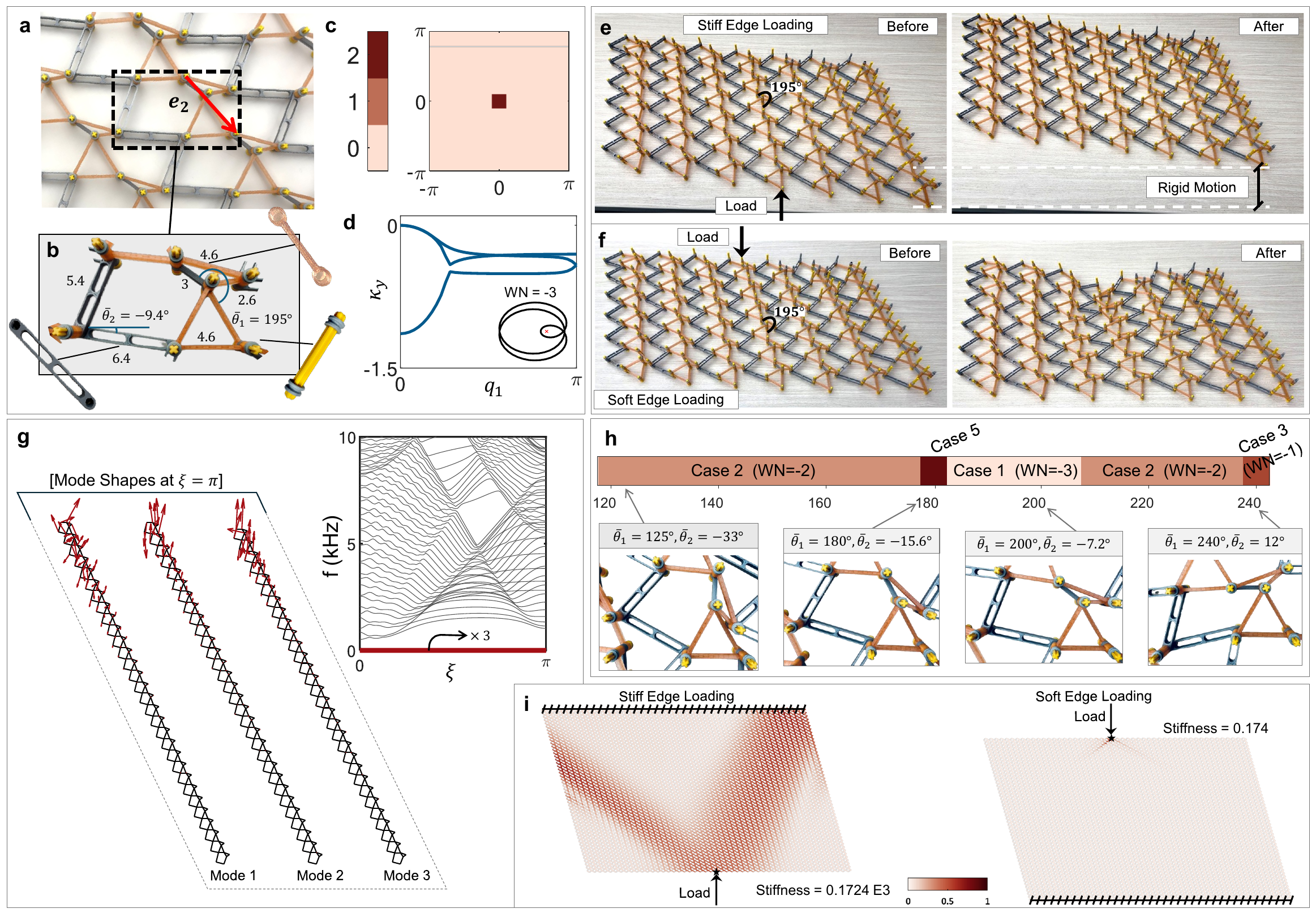}
\caption{\label{fig:6}Numerical analysis and experimental validation of the fully polarized augmented kagome lattice. (a--b) Geometric configuration of the experimental prototype and close-up of the unit cell, highlighting the 3D-printed links and the LEGO\textsuperscript{\textregistered} axles and bushings used as hinges for the selected configuration ($\bar\theta_1=195^\circ, \bar\theta_2=-9.4^\circ$). (c) Nullity of the compatibility matrix across the Brillouin zone. (d) DRs of ZMs versus wavenumber $q_1$, with the corresponding winding number indicated in the inset. (e--f) Quasi-static experimental response of a finite $6 \times 7$ prototype: loading at the stiff edge (e) induces global rigid-body motion, whereas loading at the soft edge (f) results in localized deformation (see Supplementary Video 3). (g) Band diagram of a 30-cell supercell highlighting three ZMs and their associated edge-localized mode shapes at $\xi = \pi$. (h) Evolution of the topological decay-rate diagram as a function of the configurational parameter $\bar\theta_1$ ($120^\circ \le \bar\theta_1 \le 240^\circ$), illustrating transitions between distinct Cases (1, 2, 3, and 5), with representative unit-cell geometries. (i) Computed edge stiffness and normalized displacement magnitude from full-scale static simulations of a $50 \times 50$ lattice. Probing the stiff (left) and soft (right) edges reveals an approximately three-order-of-magnitude contrast in effective edge stiffness.}
\end{figure*}

For our configuration, at each $q_1$ $\det \mathbf{C}(q_2)$ is a third-order polynomial in the Bloch factors; therefore, up to three edge ZMs may arise. Their localization is dictated by the bulk--edge correspondence, quantified by the integer-valued winding number (WN)~\cite{mao2018maxwell}:
\begin{equation}
    \nu(q_1)=\frac{1}{2\pi} \int_{-\pi}^{\pi} dq_2\frac{\partial}{\partial_{q_2}}\mathrm{Im} \log \det \mathbf{C}(q_1,q_2)
\end{equation}
This topological invariant counts the number of modes effectively transferred between opposing edges by the polarization, contributing to the excess and deficit of softness at the \textit{floppy} and \textit{stiff} edges, respectively. This calculation remains valid in the absence of bulk ZMs, i.e., when the integration contour does not intersect any zeros of $\det \mathbf{C}$. Physically, bulk ZMs (and their dual SSSs) emerge in Maxwell lattices when bonds form collinear alignments. In such cases, uniform displacements along the lines of collinearity cost no energy to linear order.

The effects of the cell geometry parameter sweep on the polarization are summarized in Fig.~\ref{fig:5}. We list ten distinct scenarios in Fig.~\ref{fig:5}b, each categorized by the number of ZMs having positive, negative, and zero DRs, and the corresponding WN. Cases 1--4 are free of bulk ZMs (modes with DR=0), whereas cases 5--10 contain at least one. The phase diagrams in Fig.~\ref{fig:5}c show the occurrence of the possible scenarios in the parameter space $(\bar{\theta}_1, \bar{\theta}_2)$ evaluated near $q_1=0$ (left) and $q_1=\pi$ (right). We note that Cases 4 and 7, which correspond to configurations in which three and two edge ZMs localize at the bottom boundary, respectively, while listed as theoretically possible scenarios in Fig.~\ref{fig:5}b, cannot be realized in our parameter space, i.e., by simply sweeping $(\bar{\theta}_1, \bar{\theta}_2)$.

Representative configurations for key scenarios (including their unit-cell geometries, compatibility matrix nullity, WNs (see SI~\ref{WNcalc} for more information regarding WN calculation), and decay-rate spectra) are discussed in Figs.~\ref{fig:5}d--e. For Cases 1--3, illustrated in Fig.~\ref{fig:5}d, the absence of collinear bond alignments precludes the existence of bulk ZMs, other than the two trivial translations at $\mathbf{q}=\mathbf{0}$, as indicated by the nullity plots. The absence of ZMs renders the polarization of these configurations particularly effective by preventing any deformation applied at the edges from leaking into the bulk. The special configuration of Case 1 (shaded green in Figs.~\ref{fig:5}b and c) has $\mathrm{WN}=-3$ and exhibits three negative DRs corresponding to three ZMs simultaneously localized at the top edge. This configuration achieves what we call \textit{full polarization}: all the modes are edge-localized, focused on the same edge and immune to bulk leakage. In Cases 2 and 3, one and two ZMs, respectively, migrate from the top boundary to the bottom boundary, as captured by the DR plots and the changes in the WN (see SI~\ref{topoe1} for topological characterization along $\mathbf{e}_1$). Fig.~\ref{fig:5}e presents representative examples of configurations featuring bulk ZMs (Cases 5--10). Here, the presence of at least one collinear structural alignment is directly linked to the emergence of lines of bulk ZMs, clearly identifiable in the respective nullity plots. Case 7 is realized here by constraining the geometry such that the constituent triangles become isosceles, with side lengths 0.57 and 0.72, while maintaining a base length of 1 (see SI~\ref{SSS} for notes on bulk ZMs and SSSs).

To demonstrate the physical manifestation of the \textit{fully polarized} phase, we construct a prototype based on Case 1 configuration from Fig.~\ref{fig:5}d. The lattice consists of slender links fabricated via 3D printing in polylactic acid (PLA) using an Original Prusa i3 MK3S $\&$ MK3S+ printer equipped with a 0.4 mm nozzle, and connected through low-friction hinges realized with LEGO\textsuperscript{\textregistered} axles and bushings. A portion of the assembled $6\times7$ lattice, with twist parameter $\bar\theta_1=195^\circ$, is shown in Fig.~\ref{fig:6}a with a detailed view of the unit cell in Fig.~\ref{fig:6}b. In this prototype, the geometry is parameterized solely by the relative angle $\bar{\theta}_1$, while $\bar{\theta}_2$ is uniquely determined by the fixed pentagonal side lengths. The nullity of the compatibility matrix (Fig.~\ref{fig:6}c) confirms the absence of bulk ZMs beyond the two trivial rigid-body translations at ($\mathbf{q}=\mathbf{0}$). The corresponding DRs (Fig.~\ref{fig:6}d) reveal three negative branches, indicating localization of ZMs at the top edge. This polarization is topologically protected by a $\mathrm{WN} = -3$ (inset, Fig.~\ref{fig:6}d). The mechanical signatures of this polarization are demonstrated through quasi-static load tests. When a load is (manually) applied at the \textit{stiff} bottom edge (Fig.~\ref{fig:6}e), the entire lattice responds through rigid-body motion. In stark contrast, loading at the \textit{soft} top edge (Fig.~\ref{fig:6}f) causes the lattice to readily deform through a localized mechanism, marking a pronounced asymmetry in the mechanical response. This motion eventually enters a geometrically nonlinear regime involving macroscopic rotations of the triangular elements, which decays rapidly into the bulk.

We further confirm the spectral signature of these edge states via supercell analysis of a domain comprising 30 cells along $\mathbf{e}_2$ (Fig.~\ref{fig:6}g). The resulting dispersion diagram features three overlapping zero-energy branches, whose corresponding mode shapes at $\xi = \pi$ are all localized at the top boundary. The arrow lengths indicate relative displacement amplitudes. To explore the sensitivity of this topological phase to geometric variations, we perform a sweep of the twist angle $\bar{\theta}_1$ over the range $120^\circ \le \bar{\theta}_1 \le 240^\circ$ (Fig.~\ref{fig:6}h). The constraints of the prototype allow for physical realization of Cases 1, 2, 3, and 5, shown in the insets (see also the Supplementary Video 3 illustrating the asymmetric edge response for three representative configurations from Cases 1--3). We note that the fully polarized regime is realized for $180^\circ < \bar\theta_1 \le 205^\circ$. The lower bound ($\bar{\theta}_1 = 180^\circ$) corresponds to Case 5, where alignment of structural elements produces collinear bonds, signaling a topological phase transition.

Finally, we quantify the asymmetry of the static response through full-scale simulations of a $50\times50$ lattice. We define the effective edge stiffness as the ratio of an applied unit-amplitude point force to the resulting displacement at the loading site (Fig.~\ref{fig:6}i)~\cite{chararaPRB2023AUGMENT}. The results reveal that the stiffness at the stiff and soft boundaries differ by nearly three orders of magnitude. The normalized displacement fields (Fig.~\ref{fig:6}i) visually confirm this dichotomy; loading at the stiff edge distributes the deformation throughout the lattice bulk, whereas loading at the soft edge produces highly confined displacement.

\section{Conclusion}
In this work, we present a systematic framework for the design and realization of polarized structural metamaterials by integrating TO algorithms with the analytical theory of topological Maxwell lattices. This approach leads to the identification of a structural geometry exhibiting a pronounced asymmetric response that remains robust under variations in boundary truncation. The predicted behavior is validated through numerical simulations and laser vibrometry measurements on a manufactured prototype. These results demonstrate that the observed edge asymmetry is a fundamental property of the bulk that manifests at the lattice boundary, rather than a consequence of specific edge features. In this respect, this robustness can be seen as functionally equivalent to the topological protection enjoyed by polarization in Maxwell lattices, albeit not describable by any formal topological invariants. We recognize that achieving such robustness is not automatically guaranteed by the satisfaction of the requirements embedded in the TO algorithm and must be verified a posteriori on a case-to-case basis. In essence, we show that, to design polarized structural metamaterials, we need an integrated strategy that combines TO with a posteriori phenomenological performance checks. 

The geometry obtained via TO is further mapped onto an ideal augmented kagome lattice amenable to topological polarization. Within this framework, we are able to identify -- and switch between -- multiple configurations with different degrees of edge asymmetry, including a \textit{fully polarized} phase with winding number $\mathrm{WN} = -3$ that supports three independent edge-localized zero modes at the soft boundary. The distinctive feature of this work is the confluence of two design perspectives: on one hand, TO and its ability to handle geometric complexity; on the other hand, cell augmentation via heuristic parameter selections, which taps into the reliable properties of kagome lattices, for which formal polarization analysis can be conducted. These two approaches are not merely used in parallel. Our contribution here is to use them \textit{jointly} in a mutually inspiring way. The properties of polarized Maxwell lattices inspire the objectives and constraints used in the TO model. The peculiar geometry obtained via TO leads to the generation of a new family of Maxwell lattices amenable to polarization, which would otherwise have been difficult to extrapolate from the kagome lattice. From a scientific standpoint, the interplay of these two perspectives -- and the unexpected insight that this led to -- is the most tangible legacy of our investigation.

\section{Acknowledgment}
P.A. acknowledges the support of the UMN Doctoral Dissertation Fellowship. P.A. and S.G. acknowledge support from the National Science Foundation (NSF grant CMMI-2344257). X.S.Z. and R.D.K. acknowledge support from the Air Force Office of Scientific Research (award FA9550-23-1-0297) and National Science Foundation (grants CMMI-2344258 and CMMI-2505649). P.A. and S.G. are grateful to M. Charara for invaluable support through his computational and experimental framework.

\bibliographystyle{apsrev4-1}
\bibliography{Ref}

\onecolumngrid
\newpage
%%%%%%%%%%%%%%%%%%%%%%%%%%%%%%%%%%%%%%
%%   Supplementary Information
%%%%%%%%%%%%%%%%%%%%%%%%%%%%%%%%%%%%%%

\makeatletter

\renewcommand\thetable{S\@arabic\c@table}
\renewcommand{\thefigure}{S\arabic{figure}}
\renewcommand \theequation{S\@arabic\c@equation}
\makeatother
\setcounter{equation}{0}  %  this will re-count eq from 1
\setcounter{figure}{0}  %  this will re-count eq from 1
\setcounter{section}{0}  %  this will re-count eq from 1

\maketitle
\section*{\Large Boosting lattice polarization \\ Mixing the perspectives of geometry optimization and cell-augmentation: Supplementary Information}

\section{Partially Polarized and Non-Robust Optimized Designs\label{addl_TO_designs}}

\begin{figure}[h!]
\includegraphics[width=\columnwidth]{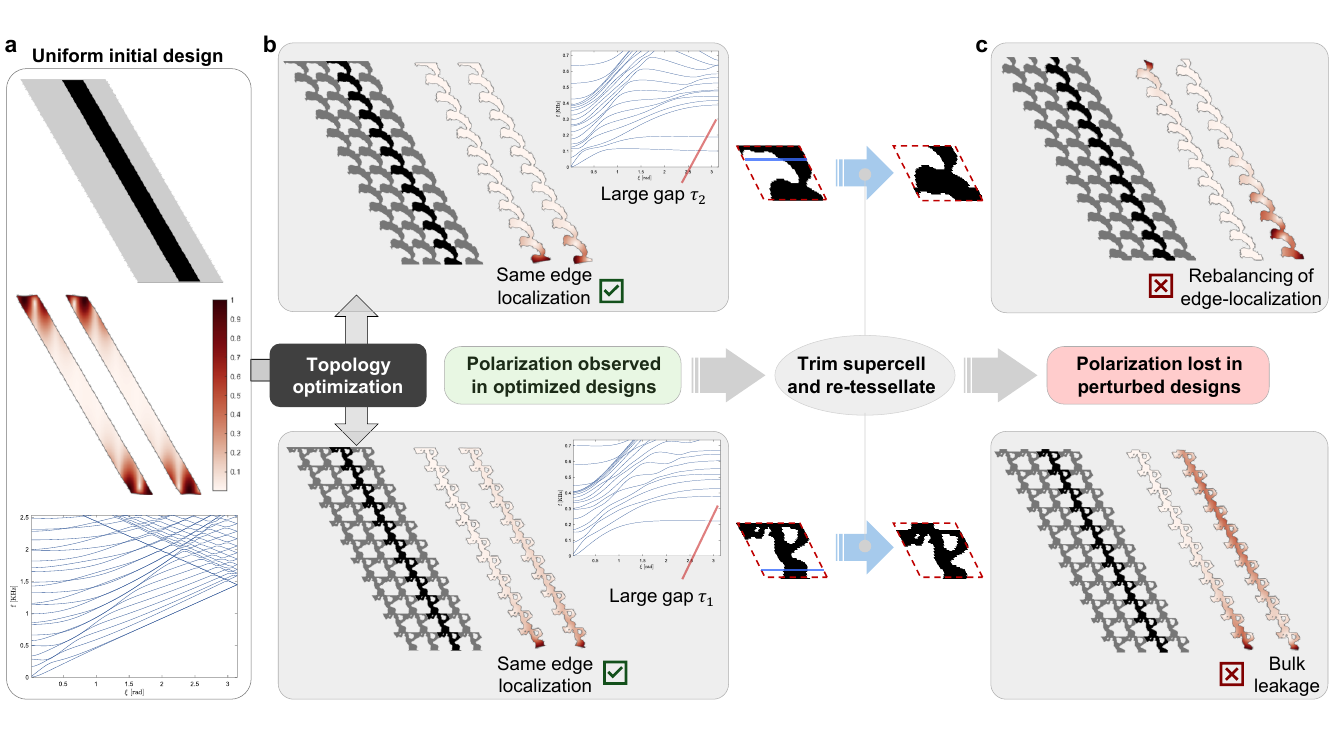}
\caption{Examples of non-robust polarization in topology-optimized designs. (a) Uniform initial design used as the starting point for topology optimization. (b) Two representative optimized configurations exhibiting apparent polarization, characterized by localization of the two lowest-frequency modes at the same edge and a pronounced separation from the remaining spectrum. (c) Robustness assessment performed by trimming the optimized supercells and re-tessellating the resulting geometries. In both examples, the polarization is lost after perturbation: the upper case exhibits a redistribution of localization between opposite edges, whereas the lower case develops bulk leakage. These results demonstrate that satisfying the optimization objectives alone does not guarantee polarization that is robust to boundary modifications.}\label{fig:S1_a}
\end{figure}

\begin{figure}[h!]
\includegraphics[width=\columnwidth]{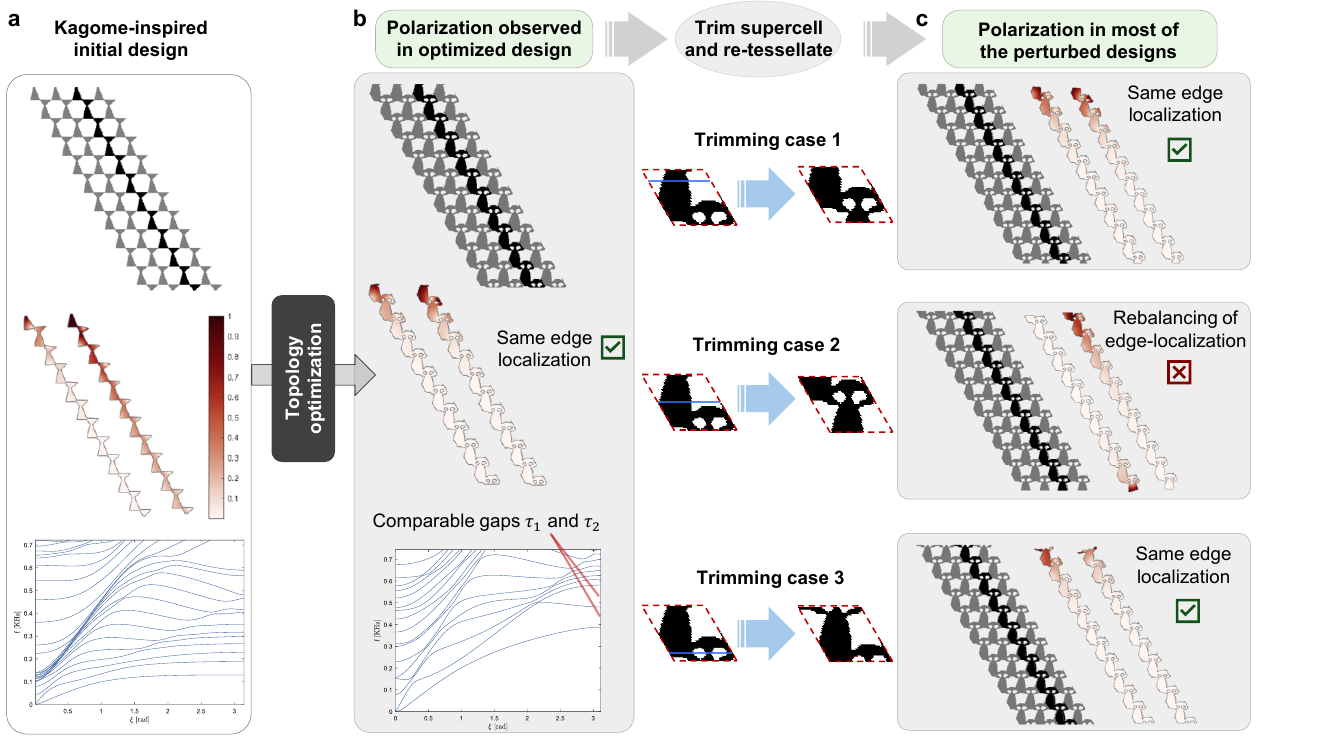}
\caption{Effect of an informed initial guess on optimization. (a) Kagome-inspired initial guess used for topology optimization. (b) The optimized design exhibiting stronger polarization characteristics, with increased isolation of the first two lowest-frequency modes and enhanced same-edge localization of modal energies. (c) Representative trimming scenarios showing that polarization is preserved in most cases, although certain trims may still lead to a redistribution of edge localization.}\label{fig:S1_b}
\end{figure}

In our experience, obtaining a design whose polarization originates as a bulk property is challenging. With the optimization formulation described in sec.~\ref{sec:TO_framework}, the optimizer successfully finds a variety of designs that satisfy the desired primary criteria, i.e., same-edge localization of the first two dominant modes. However, most of these designs do not exhibit robust polarization against different boundary truncations, as these criteria are not embedded in the optimization formulation, primarily owing to the significant computational cost associated with repeated eigenvalue problem solutions and eigenvector sensitivity computations. Fig.~\ref{fig:S1_a} includes two such representative examples where the optimized design is polarized in its unperturbed form, but the polarization property is lost when subjected to boundary trimming. Here, the optimization starts from a uniform initial guess (i.e., $z_e=0.5, e=1,\ldots,N_e$) (Fig.~\ref{fig:S1_a}a) and produces two optimized designs that show both band isolation and same-edge localization of modal energies for the first two modes (Fig.~\ref{fig:S1_a}b). However, the polarization properties are lost under small geometric perturbations introduced through trimming and re-tessellation, as shown in Fig.~\ref{fig:S1_a}c with representative examples for both optimized designs corresponding to different trimming heights. The optimized design in the top row of Fig.~\ref{fig:S1_a}b supports strong isolation of the first two bands from the rest of the spectrum at $\xi=\pi$, and the corresponding mode shapes display pronounced same-edge localization. However, inspection of the mode shapes reveals that this behavior depends strongly on the specific free-edge geometry, which includes a nearly dangling material block connected with the remainder of the design through a thin hinge-like member. As a result, the polarization property is lost when the design is trimmed at a different height and these edge features are perturbed. For the optimized design shown in the bottom row of Fig.~\ref{fig:S1_a}b, we observe a large gap between the first and second bands at $\xi=\pi$, together with strong localization of modal energies at the same edge. In this case, perturbation of the geometry transforms the second mode from an edge mode into a bulk mode, as evidenced by the bulk leakage observed in the mode shape of the second mode of the perturbed design.  

To improve the situation, one could adopt a robust optimization formulation that explicitly accounts for boundary trimmings at different heights by solving additional eigenvalue problems on the trimmed configurations and enforcing the polarization requirements for each trimming scenario directly within the optimization process. However, the large number of eigenvalue problems and associated sensitivity analysis result in a significant computational burden. Alternatively, informed initial guesses can be used to increase the likelihood of achieving robust polarization in the optimized design. Fig~\ref{fig:S1_b} shows one such example, where a Kagome-inspired initial guess (Fig~\ref{fig:S1_b}a) is used to obtain an optimized design with improved polarization characteristics. Specifically, the optimized design exhibits increased isolation of the two lowest bands at $\xi=\pi$ together with stronger same-edge localization of modal energies (Fig.~\ref{fig:S1_b}b). Furthermore, these polarization properties are retained for most trimming configurations obtained by cutting the design at different heights. Fig~\ref{fig:S1_b}c presents two representative cases in which the polarized behavior is preserved, together with one case where a rebalancing of edge location occurs. 

\section{The 3D Laser Doppler Vibrometer Experiments\label{laserexperiment}}

\begin{figure}[h!]
\includegraphics[width=0.9\columnwidth]{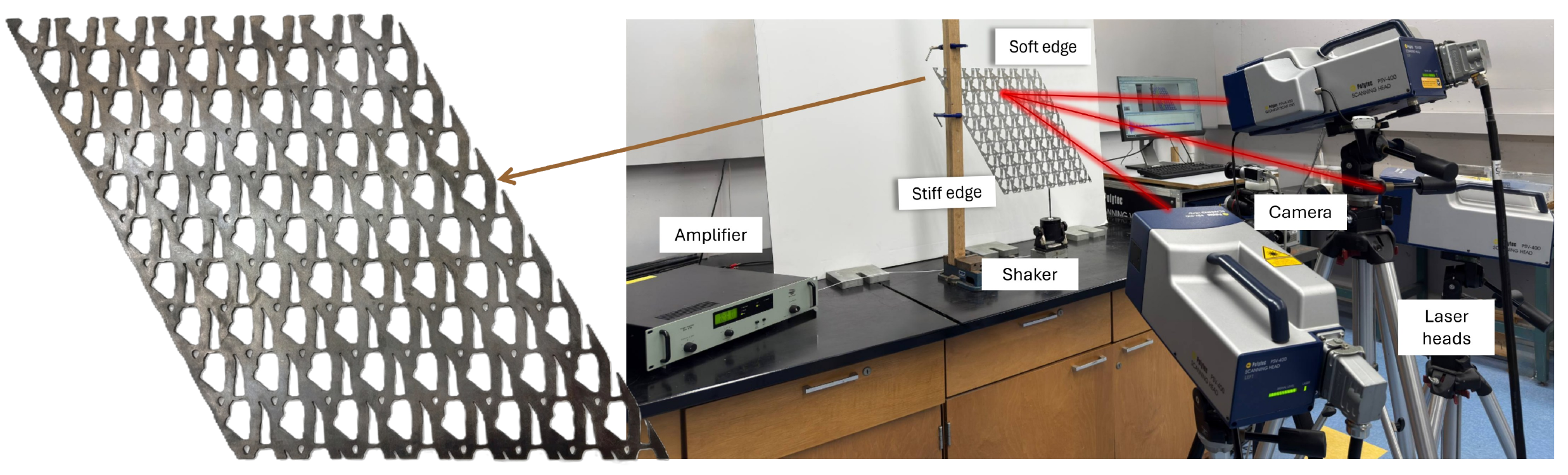}
\caption{\label{fig:S2}Experimental setup for the 3D scanning laser Doppler vibrometer measurements performed on the waterjet-cut lattice specimen derived from the topology optimization results. A close-up view of the specimen is shown on the left.}
\end{figure}

The topology optimization process yields a pixelated black-and-white 11-cell supercell (Fig.~\ref{fig:1}b), where black and white regions correspond to solid and void phases, respectively. Owing to the jagged nature of the optimized boundaries, the unit-cell geometry is imported into SolidWorks, where spline-based smoothing is applied to regularize the boundaries. The refined geometry is subsequently scaled such that the side length of each unit cell equals 5~cm. For the supercell analysis presented in Fig.~\ref{fig:2}a, the smoothed geometry is tessellated to construct the finite supercell domain and exported as a \texttt{.STEP} file for meshing in GMSH. The domain is discretized using four-node isoparametric quadrilateral elements under a two-dimensional plane-stress assumption. Each node possesses two in-plane translational degrees of freedom, and a unit thickness is assumed throughout the simulations. The material is modeled as linearly elastic and isotropic. Standard $2\times2$ Gauss quadrature is employed for numerical integration during assembly of the global stiffness and mass matrices. Free boundary conditions are prescribed at the top and bottom edges of the supercell, while Bloch-periodic boundary conditions are applied in the transverse direction. The resulting eigenvalue problem is then solved to obtain the supercell dispersion relations. 

For full-scale transient simulations, the refined unit cell is tessellated into a finite $10\times8$ lattice domain. Throughout the simulations we assume free boundary condition. The same geometry is additionally exported as a \texttt{.SLDPRT} file for fabrication of the experimental specimen. The manufactured lattice, shown in the left panel of Fig.~\ref{fig:S2}, is produced via waterjet cutting from a 2-mm-thick aluminum sheet. The aluminum material properties used in the numerical simulations are Young’s modulus $E=71$~GPa, Poisson’s ratio $\nu=0.33$, and mass density $\rho = 2700~\mathrm{kg/m^3}$.

Experimental characterization of wave propagation is performed using a Polytec PSV-400 3D Scanning Laser Doppler Vibrometer (SLDV) equipped with three scanning laser heads for full in-plane velocity measurements. The excitation signal is first generated in MATLAB and subsequently imported into the PSV software as a user-defined excitation waveform. The signal is routed through a loop-back BNC connection (Generator Out to Ref 1) to a power amplifier (Brüel \& Kj\ae r Type 2718), which drives an electromechanical shaker (Brüel \& Kj\ae r Type 4810). The specimen is excited through an attached stinger using super glue at the loading location. The measured velocity fields from the three laser heads are internally transformed into the $\hat{x}$, $\hat{y}$, and $\hat{z}$ components through the PSV 3D alignment procedure based on Euler-angle decomposition.

To investigate the asymmetry of wave transport between the soft and stiff boundaries, three sets of transient measurements are conducted using a 9-cycle tone burst excitation with a carrier frequency of 7.7~kHz. In the first setup, the specimen is mounted such that the stiff edge is positioned at the bottom boundary. The left edge of the lattice is clamped between two wooden supports, while the remaining boundaries are left free. Excitation is then applied at the center of the stiff boundary, as illustrated in Fig.~\ref{fig:S2}. The specimen is then remounted to position the soft boundary at the bottom, and the experiment is repeated for two distinct loading conditions, the outer soft edge and the inner soft edge, to match the numerical cases analyzed in the main text (Fig.~\ref{fig:3}c).

We record the measurements at a sampling frequency of 51.2~kHz, with a spatial resolution of four scan points per unit cell to ensure sufficient data for wavefield reconstruction. Retroreflective tape is applied at each scan location to improve optical return quality and reduce measurement noise. Each test sequence automatically moves the lasers to subsequent scan locations. We provide sufficient relaxation time between scans to ensure that bursts fully dissipate by damping before the next measurement. To mitigate random background noise, each measurement is averaged over 15 readings, retaining only the components synchronized with the trigger. We use an internal trigger with a 1\% pre-trigger to record data immediately preceding the burst, ensuring the propagating wavefront onset is clearly captured. In addition, a built-in high-pass digital filter with a cutoff frequency of 1~kHz is applied within the PSV software to suppress low-frequency environmental vibrations.

Post-processing of the measured velocity fields is performed in MATLAB to reconstruct the transient wave propagation. A zero-phase digital bandpass filter (5.7--9.7~kHz) is first applied to the raw in-plane velocity components, $v_x$ and $v_y$, to suppress spurious frequency contributions. The analytic signal associated with each measurement point is then constructed using a Hilbert transform, allowing extraction of the instantaneous amplitude of the wavefield. The resultant in-plane energy envelope is then calculated to visualize the propagating wave without the interference of carrier-wave phase oscillations or cyclic zero-crossings. This processing significantly enhances the signal-to-noise ratio, providing a clear mapping of wave localization and the contrast in energy transport.

\section{Remarks about Winding Number Calculation\label{WNcalc}}

We recall that the topological winding number (WN) is defined by the phase of $\det \mathbf{C}(\mathbf{q})$ as the wavevector $\mathbf{q}$ traverses the cycles of the Brillouin zone. This WN is path-independent and remains constant so long as the phonon spectrum is gapped everywhere except at the origin ($\mathbf{q}=\mathbf{0}$)~\cite{lubensky2015phonons}. While the zero mode (ZM) at $\mathbf{q}=\mathbf{0}$ is not topologically protected and thus does not affect the global invariant, certain configurations may host topologically protected bulk ZMs at other wavevectors. These modes, around which the phase of $\det \mathbf{C}(\mathbf{q})$ advances by $2\pi$, serve as mechanical analogs to Dirac semimetals in electronic systems~\cite{rocklin2016mechanical}. 

\begin{figure}[h!]
\includegraphics[width=0.9\columnwidth]{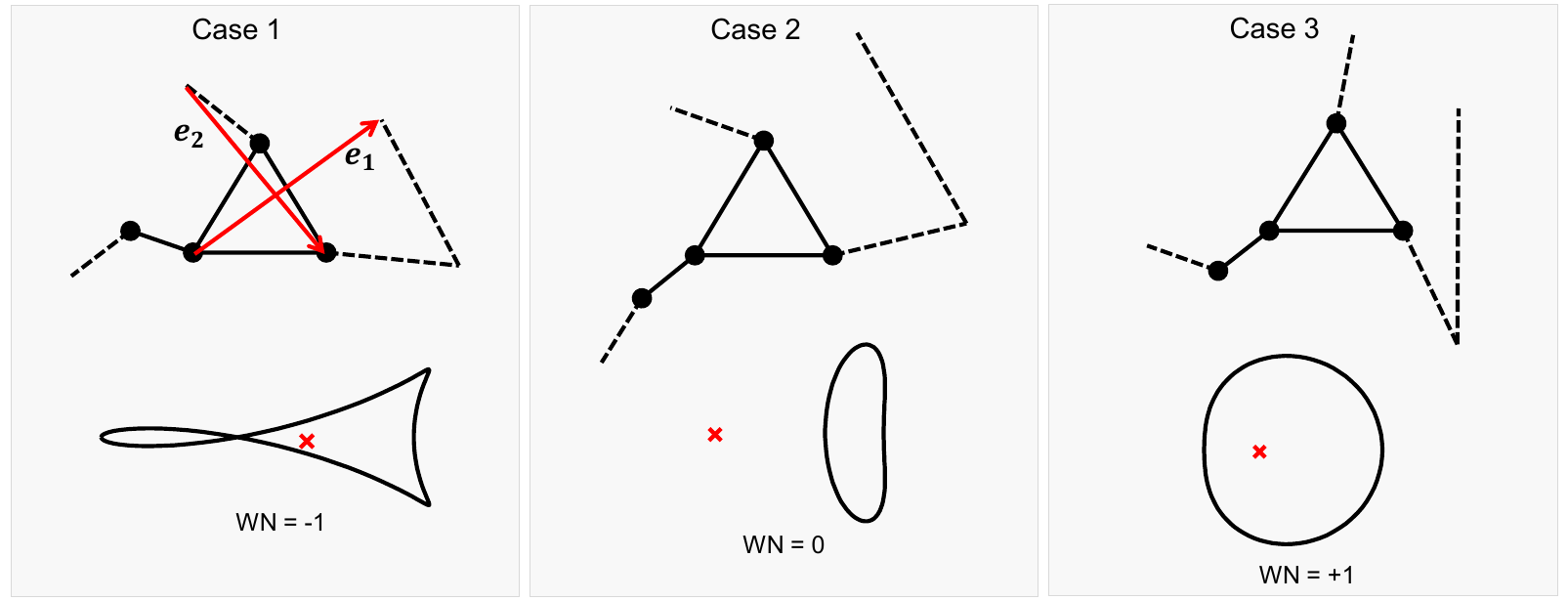}
\caption{Alternative gauge choices and corresponding WNs for Cases 1--3. Above each panel, the unit-cell geometry is illustrated, where solid lines represent intracell bonds and dashed lines denote intercell connections to adjacent units. While the absolute WNs ($\mathrm{WN} = -1, 0, +1$) differ from the main text due to the choice of gauge, the relative increment remains consistent, capturing the migration of zero-energy modes between boundaries.\label{fig:S3}}
\end{figure}

It is important to note that the WN is not gauge-invariant; it depends explicitly on the choice of the reference unit cell. However, by maintaining a consistent gauge (i.e., adopting the same unit cell definition across different configurations), we can reliably track the migration of edge-localized ZMs between opposing boundaries. In the main text, we utilize the primary unit cell shown in Fig.~\ref{fig:5}a. For Cases 1, 2, and 3, this choice yields $\mathrm{WN}= -3$, $-2$, and $-1$, respectively. This sequence quantifies the spatial transition: in Case 1, three ZMs are localized at the top boundary (indicated by three negative decay rates). As we move to Case 2 and then Case 3, we observe the migration of these modes to the bottom boundary, corresponding to the incremental change in the WN and the appearance of positive decay rates in the spectrum. Another choice of conventional unit cell is a symmetric unit cell choice where all bonds and sites are centered at the cell's origin~\cite{mao2018maxwell}. This alternative is shown in Fig.~\ref{fig:S3} for Cases 1–3. While this choice shifts the absolute WNs to $\mathrm{WN} = -1$, $0$, and $+1$, respectively, the relative change between Cases remains identical ($\Delta\mathrm{WN} = +1$ per step). Notably, this change in unit cell definition does not alter the primitive lattice vectors $\mathbf{e}_1$ and $\mathbf{e}_2$, nor does it affect the decay rate diagrams.

\section{Topological Characterization along $\mathbf{e}_1$ Direction\label{topoe1}}

\begin{figure}[h!]
\includegraphics[width=0.9\columnwidth]{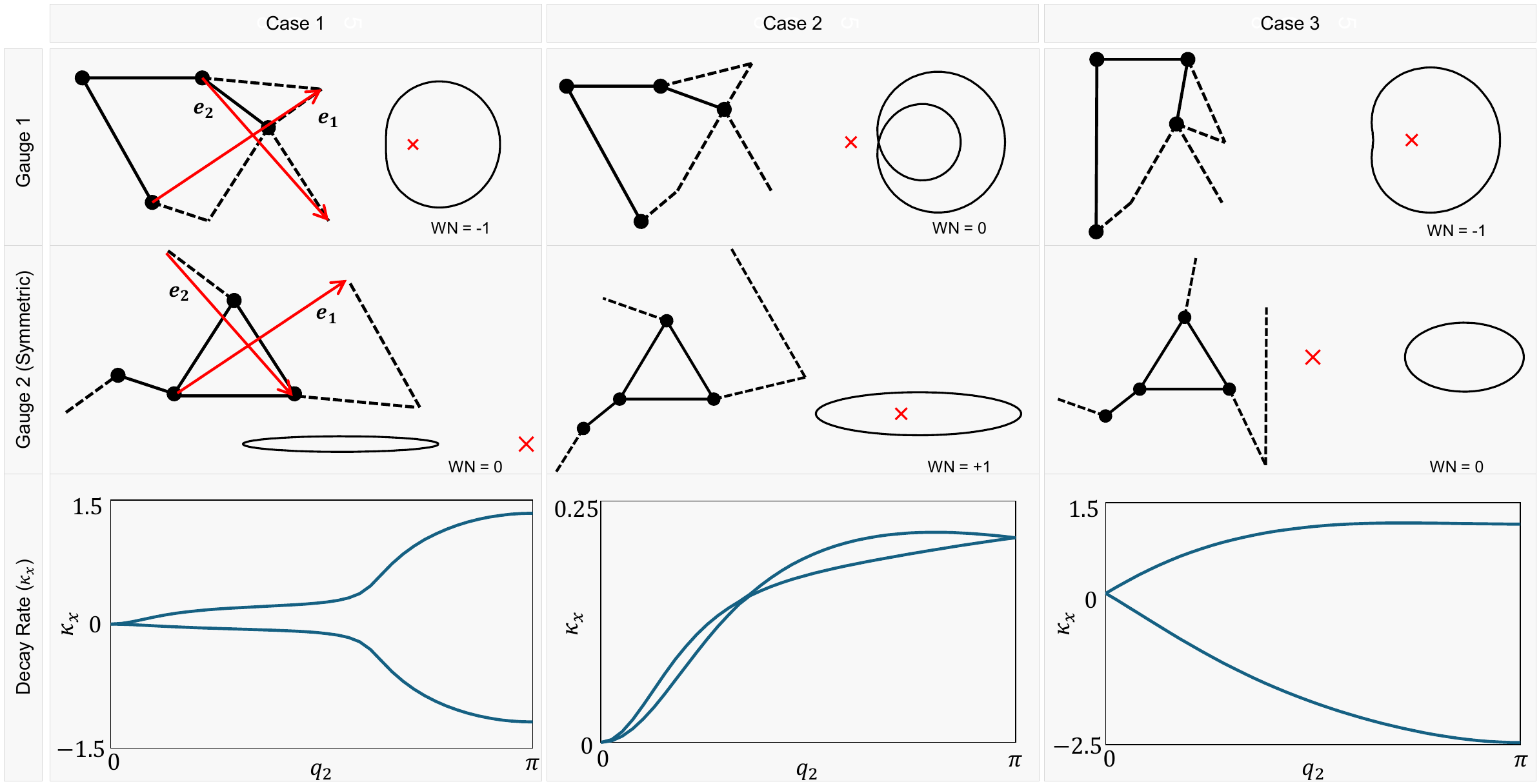}
\caption{Decay rates $(\kappa_x)$ and WN $(\nu(q_2))$ for Cases 1--3 calculated for two different gauge choices.\label{fig:S4}}
\end{figure}
The polarization characterization presented in the main text for Cases 1--3 is performed for lattice truncations along the $\mathbf{e}_2$ direction. Specifically, we analyze the decay rates $\kappa_y(q_1)$, which govern the exponential localization of ZMs at the edges along $\mathbf{e}_2$. In that configuration, Case 1 supports three ZMs localized at the negative $\mathbf{e}_2$ boundary (top edge). In Cases 2 and 3, these modes sequentially migrate to the opposite boundary, resulting in distributions of $(2, 1)$ and $(1, 2)$ modes at the negative and positive $\mathbf{e}_2$ boundaries, respectively. This transition is captured by the WN $\nu(q_1)$, which follow the sequence $-3, -2, -1$ for the primary unit cell (Fig.~\ref{fig:5}a) and $-1, 0, +1$ for the symmetric gauge (Fig.~\ref{fig:S3}). To provide a complete topological profile, we repeat this analysis for edges oriented along the $\mathbf{e}_1$ direction by evaluating the decay rates $\kappa_x(q_2)$ and the corresponding WNs $\nu(q_2)$. The results for Cases 1--3 are illustrated in Fig.~\ref{fig:S4} for both gauge choices. A key distinction arises here: whereas $\det \mathbf{C}(q_1, q_2)$ is a third-order polynomial in the Bloch factors for a fixed $q_1$, it reduces to a second-order polynomial for a fixed $q_2$. Consequently, the system supports a maximum of two distinct edge modes along the $\mathbf{e}_1$ direction, rather than three.

As shown in Fig.~\ref{fig:S4}, Cases 1 and 3 exhibit one positive and one negative decay rate $\kappa_x$, indicating that one ZM is localized at each opposing boundary. In the symmetric gauge, this \textit{balanced} distribution yields a WN of $\nu(q_2) = 0$, signifying that these configurations are effectively unpolarized along $\mathbf{e}_1$. In contrast, Case 2 possesses two positive decay rates, meaning both ZMs are localized at the boundary pointing toward the positive $\mathbf{e}_1$ direction. This polarized state is reflected by a WN of $\nu(q_2) = +1$ in the symmetric gauge. These findings demonstrate that, while the optimized lattices are designed for extreme asymmetry along a specific axis, their topological response is inherently anisotropic, with the degree of polarization governed by the specific orientation of the lattice truncation.

\begin{figure}[h!]
\includegraphics[width=0.8\columnwidth]{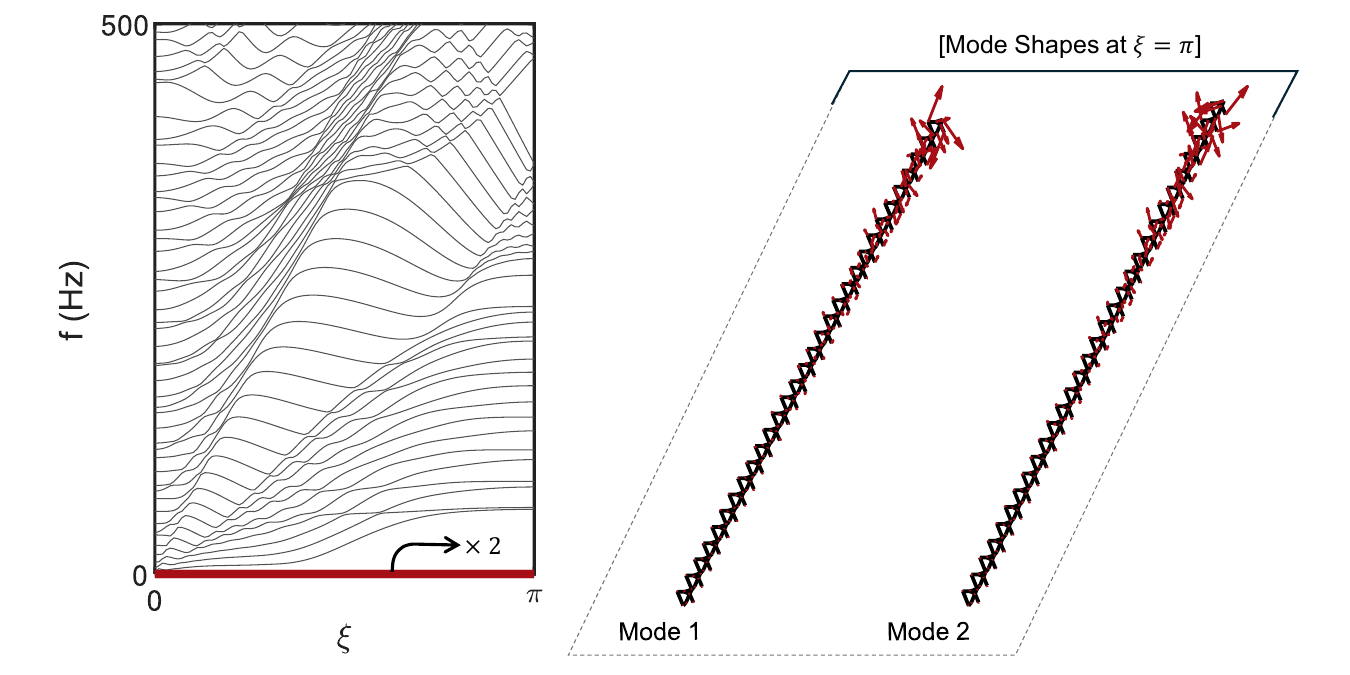}
\caption{Supercell eigenmode analysis for Case 2 along the $\mathbf{e}_1$ direction. (Left) Band diagram for a 30-cell supercell (finite along $\mathbf{e}_1$, periodic along $\mathbf{e}_2$), highlighting two zero-frequency edge modes (red). (Right) Mode shapes for the two ZMs at $\xi = \pi$, showing clear localization at the positive $\mathbf{e}_1$ boundary, consistent with the predicted WN.\label{fig:S5}}
\end{figure}

To further corroborate the topological polarization of Case 2 along the $\mathbf{e}_1$ direction, we perform a supercell analysis to visualize the localization of the zero-frequency modes. We construct a supercell comprising 30 unit cells of Case 2 geometry, applying Bloch periodic boundary conditions along the $\mathbf{e}_2$ direction while maintaining free boundaries along $\mathbf{e}_1$. Fig.~\ref{fig:S5} presents the resulting band diagram, which features two zero-frequency modes (highlighted in red). The corresponding mode shapes, evaluated at $\xi = \pi$, clearly demonstrate that these ZMs are localized at the boundary oriented toward the positive $\mathbf{e}_1$ direction. This numerical result is in agreement with the calculated WN ($\nu(q_2) = +1$) and the existence of two positive decay rates ($\kappa_x(q_2)$), providing evidence of the lattice's directional polarization.

\section{Note on Bulk Zero-Energy Modes\label{SSS}}

\begin{figure}[h!]
\includegraphics[width=0.9\columnwidth]{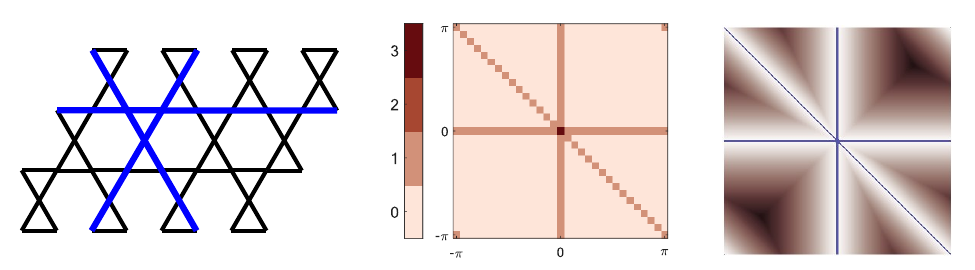}
\caption{Reference analysis for the regular kagome lattice. From left to right: Lattice geometry with the three primary SSSs highlighted in blue; nullity map of the compatibility matrix across the Brillouin zone, showing three intersecting lines of mechanisms; dispersion surface of the first mode, where zero-frequency branches (blue lines) align with the SSS directions predicted by the nullity plot. \label{fig:S6}}
\end{figure}

The regular kagome lattice is characterized by three primary directions, rotated by $60^\circ$, along which the bonds align perfectly to form continuous, load-carrying fibers. These collinear structures support states of self stresses (SSSs), which are fundamentally linked to the existence of Bulk ZMs. In the phonon spectrum, these bulk ZMs manifest as flat, zero-frequency branches for wavevectors along the $\Gamma-\mathrm{M}$ directions of the Brillouin zone, corresponding to the three fiber orientations (Fig.~\ref{fig:S6})~\cite{Azizi-Gonella_Kagome-chain_PRApp_2024}. Deviating from this regular configuration (e.g., in a twisted kagome) breaks these collinear fibers, thereby lifting the zero-frequency branches to finite frequencies across the entire Brillouin zone, with the exception of the origin.

\begin{figure}[h!]
\includegraphics[width=\columnwidth]{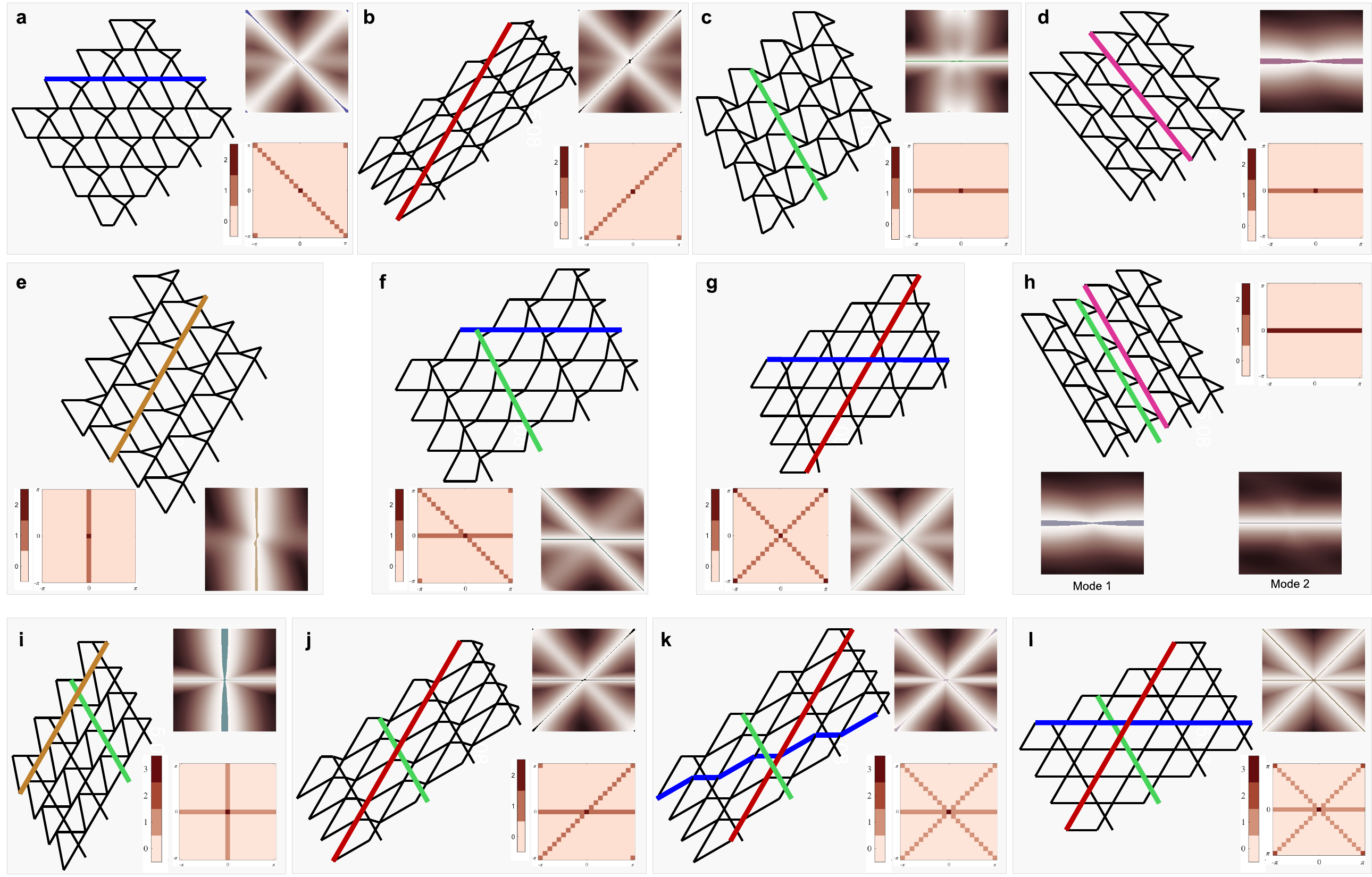}
\caption{Gallery of augmented kagome configurations with diverse bulk ZM signatures. For each panel (a–l), we display the lattice geometry with color-coded SSS fibers, the nullity map of the compatibility matrix, and the corresponding dispersion surface for the first phonon mode with zero frequency branches highlighted. Panel (h) presents the dispersion surfaces of the first two modes. These Cases demonstrate the ability to program both the count and orientation of bulk mechanisms by modifying the underlying unit cell topology.\label{fig:S7}}
\end{figure}

In the augmented kagome framework introduced in this work, the expanded design space gives rise to a substantially richer variety of collinear configurations and associated bulk mechanisms. Fig.~\ref{fig:S7} presents 12 representative lattice realizations derived from this framework, each characterized by distinct lines of SSSs and their corresponding bulk ZMs. Each panel includes the lattice geometry with collinear fibers highlighted, the nullity map of the compatibility matrix, and the associated dispersion surfaces, with zero-frequency branches emphasized. Cases a--e exhibit a single SSS fiber, resulting in a nullity of one along the corresponding direction in reciprocal space. The orientations of these fibers vary (for example, Cases C and D possess similar alignments) with each configuration supports only one independent bulk mechanism. However, by combining multiple SSS orientations within a single unit cell, we can achieve configurations with an increased number of mechanisms. Cases f--l display multiple bulk-ZM directions, as evidenced by the intersecting lines in their nullity and dispersion plots. All configurations possess two trivial ZMs in the center of the Brillouin zone, except for Cases k and l, which exhibit three ZMs at $\mathbf{q}=\mathbf{0}$. Notably, specific combinations of collinear fibers, such as the combination of the patterns observed in Cases c and d, produce configurations in which the first two dispersion surfaces exhibit zero frequency lines along the $(-\pi,0)$ to $(\pi,0)$ direction. Furthermore, Cases g and l support two bulk ZMs located at the corners of the Brillouin zone.

\end{document}